\documentclass[12pt]{article}
\usepackage[letterpaper, margin=1in]{geometry}
\usepackage{amsmath}
\usepackage[noend]{algpseudocode}
\usepackage{algorithm}
\usepackage{amssymb}
\usepackage{graphicx}
\usepackage{psfrag}
\usepackage{xurl}
\usepackage{hyperref}
\usepackage{enumitem}
\usepackage{booktabs}
\usepackage{natbib}
\usepackage{xcolor}
\usepackage{subcaption}

\newcommand{\reals}{{\mbox{\bf R}}}

\newcommand{\diag}{\mathop{\bf diag{}}}

\newcommand{\ie}{{\it i.e.}}
\newcommand{\eg}{{\it e.g.}}

\newcommand{\ones}{\mathbf 1}

\newcommand{\BIT}{\begin{itemize}}
\newcommand{\EIT}{\end{itemize}}
\newcommand{\BEQ}{\begin{equation}}
\newcommand{\EEQ}{\end{equation}}
\newcommand{\BEAS}{\begin{eqnarray*}}
\newcommand{\EEAS}{\end{eqnarray*}}

\newcounter{algorithmctr}[section]
\renewcommand{\thealgorithmctr}{\thesection.\arabic{algorithmctr}}
{\refstepcounter{algorithmctr}
\begin{list}{}{%
\setlength{\rightmargin}{0\linewidth}%
\setlength{\leftmargin}{.05\linewidth}}%
\rmfamily\small
\item[]{\setlength{\parskip}{0ex}\hrulefill\par%
\nopagebreak{\bfseries\textsf{Algorithm \thealgorithmctr~}}}}%
{{\setlength{\parskip}{-1ex}\nopagebreak\par\hrulefill}
\end{list}}

{\refstepcounter{algorithmctr}
\begin{list}{}{%
\setlength{\rightmargin}{0\linewidth}%
\setlength{\leftmargin}{.05\linewidth}}%
\rmfamily\small
\item[]{\setlength{\parskip}{0ex}\hrulefill\par%
\nopagebreak{\bfseries\textsf{Algorithm}}}}%
{{\setlength{\parskip}{-1ex}\nopagebreak\par\hrulefill}
\end{list}}

\begin{document}

\title{Simple Dynamic Stock/Bond/Gold Portfolios}
\author{Nikhil Devanathan \and Alexandros E. Tzikas \and Stephen P. Boyd}
\date{\today}
\maketitle

\begin{abstract}
For more than four decades, the 60/40 stock/bond portfolio has served as a
benchmark for delivering reasonable returns without excessive risk.
More recently, a 50/30/20 stock/bond/alternative portfolio has been suggested.
We use gold as the alternative and as an inflation hedge.
In this paper we ask: how much improvement over these benchmark fixed-weight
portfolios
can be obtained using widely available public data and
standard methods from quantitative finance?

We restrict ourselves to long-only dynamic portfolios of
stocks, bonds, and gold, plus cash, rebalancing monthly,
using only publicly available data.
We evaluate portfolios on the conventional metrics: return,
volatility, Sharpe ratio (computed in excess of the federal funds rate),
drawdown, and turnover, in addition to consistency of performance
over time, judged by the consistency of the realized annual volatility.

Over the 20--year period 2006--2026,
using a conservative estimate of trading
costs, we show that all risk-adjusted and drawdown metrics are improved
using simple
volatility control, where we dynamically mix the
fixed-weight portfolios with cash so as to target a fixed volatility.
This method relies on a simple estimate of portfolio volatility derived from
past returns.
We also demonstrate that more sophisticated portfolios based
on convex optimization---similar
to those used in quantitative hedge funds---yield further
substantial improvement in return and risk-adjusted return.
We consider two such portfolios, one that uses a simple estimate
of future returns based on past returns, and one that forecasts future returns
based on past returns and just a handful of widely available public
economic data.
These portfolios also outperform a suite of standard risk-based allocation
methods, such as risk parity and minimum variance, evaluated on the same
assets and data.

Our paper is accompanied by open-source
software that implements the methods and replicates all results.
\end{abstract}

\clearpage

\section{Introduction}\label{s-intro}

A risk-averse investor seeking steady growth of capital is
conventionally advised to hold a fixed-weight 60/40 portfolio of broad
stock and bond indices, rebalanced annually
\cite{Bernstein2000, Bogle2017}.
As the correlation between stock and bond
returns has risen in recent years \cite{Brixton2023},
variants that add a diversifying third
asset class have gained traction, such as a 50/30/20
stock/bond/alternative portfolio \cite{Fink2025}.
Motivated by this proposal and by proposals to use gold as an inflation hedge
\cite{Yasmin2025}, we use a 50/30/20 stock/bond/gold portfolio as a
gold-augmented benchmark.

We ask a simple question: how much better than these fixed-weight portfolios can we do using
standard tools of quantitative finance?
We work strictly within the constraints that
make this simple setting attractive to an individual
investor: only the three broad assets (stocks, bonds, and gold),
monthly rebalancing, long only, no leverage, no derivatives,
and no proprietary data.
We restrict ourselves to methods
that are simple and completely transparent,
can be implemented in just a few lines of code,
and rely only on publicly available data.

Given these constraints, and the very small set of assets we invest in,
we do not expect hedge-fund-level results.
Instead we judge our methods against the classical 60/40 portfolio and
our 50/30/20 gold-augmented benchmark.
We find that we can obtain a substantial lift in performance
compared to these benchmarks.

We state our contribution plainly.
It is not methodological: the methods we use---volatility targeting, momentum, regression-based return
forecasting, and mean-variance optimization with a risk constraint---are all
standard and well established.
What is new is empirical: what these standard components deliver under the
constraints an individual investor actually faces---three liquid assets plus
cash, long only, no leverage, monthly rebalancing, only public data, and net
of trading costs---measured against the 60/40 and our 50/30/20 benchmark.
We evaluate them not only on the usual metrics, but also on an
inflation-adjusted and a post-tax basis.
Finally, we give a transparent and fully reproducible implementation, with
open-source code, that a self-directed investor could run themselves.

We show this using the exchange-traded funds (ETFs) SPY
(broad U.S.~equity), AGG (U.S. investment-grade bond market), and GLD (gold),
plus cash, with simulations over the 20--year period 2006--2026,
net of a conservative estimate of trading costs.
We compare three families of portfolios.
\begin{itemize}
\item \emph{Fixed-weight benchmarks.}
The conventional 60/40 and our 50/30/20 baseline portfolios
holding SPY/AGG/GLD, rebalanced annually.
\item \emph{Volatility-controlled portfolios.} The same fixed-weight
portfolios, diluted with cash each month so that their
estimated (ex-ante) volatility equals a target, such as
7\% annualized \cite{Devanathan2026SingleAsset}.
\item \emph{Optimization-based portfolios.}
Monthly rebalanced portfolios found using mean-variance or, more generally, Markowitz-type
methods, which use convex optimization to maximize a forecast of forward return
subject to a limit on ex-ante volatility \cite{Boyd2024}. We consider
two return forecasts:
a simple trailing average of past returns, and
a regression forecast based on asset and economic features.
We refer to the two associated portfolios as simple Markowitz and Markowitz,
respectively.
\end{itemize}

We judge these portfolios using conventional metrics such as return,
risk, average and maximum drawdown.  Our principal objective is Sharpe ratio,
computed as the ratio of excess return (above the risk-free
federal funds rate) to the volatility.
We also consider the steadiness or consistency of performance,
judged by the variation in annual realized risk across different economic regimes.
We find that volatility control improves these metrics over
the fixed-weight benchmarks, with a substantial increase in Sharpe ratio
along with more consistent performance.
The Markowitz portfolios improve the metrics further, lifting the Sharpe ratio
to $1.08$, nearly twice the Sharpe ratio of the 60/40 portfolio ($0.56$).
We find that these improvements persist when they are judged
using inflation-adjusted or post-tax metrics.

Since both volatility control and our Markowitz portfolios size positions using
an estimate of risk, it is fair to ask whether their advantage is just a generic
benefit of risk control.
To answer this we also compare against seven standard risk-based allocation
methods---among them risk parity, equal risk contribution, minimum variance,
and Black--Litterman---using the same assets, data, costs, and constraints.
None of them reaches the performance of our Markowitz portfolio: at a common
volatility target their Sharpe ratios lie between $0.60$ and $0.85$, against
$1.08$.
We give this comparison in appendix~\ref{a-risk-based}.

Open-source software that implements all methods and reproduces
every result accompanies the paper and can be found at
\begin{center}
{\color{blue}\url{https://github.com/cvxgrp/simple-portfolio-code}}.
\end{center}

\subsection{Related work}\label{ss-prior}
The fixed-weight stock/bond portfolio is the benchmark against which we measure performance.
Its appeal rests partly on evidence that a portfolio’s long-term allocation across
broad asset classes—for example, its target weights in stocks and bonds—explains most
of the variation in that portfolio’s returns over time, whereas security selection and
short-term market-timing decisions explain substantially less \cite{Brinson1986,Brinson1991}.
At a high level, equities
supply the return and bonds hedge equity drawdowns in a portfolio.
This hedge is good when
the stock/bond correlation is negative or positive and small.
However, this correlation is regime-dependent:
negative when growth shocks dominate but positive when inflation shocks occur
\cite{Ilmanen2003, Campbell2017, Brixton2023}. Nevertheless, evidence suggests that there is an
overall diversification benefit to holding bonds \cite{laipply2025can}.
As episodes of positive stock–bond correlation have become more prominent in recent years, especially after 2010, the
diversification benefit of the 60/40 portfolio has shrunk \cite{Smith2025LPL6040}, motivating the inclusion of a
third, separately driven asset. Gold was proposed as it can both hedge equities and act as
a safe haven in market stress \cite{Hillier2006, Baur2010a, Baur2010b}. Unlike the private
alternatives typically used for the same purpose, gold is directly accessible to a self-directed
investor. There are, however, counterarguments to including gold in a portfolio, as researchers have questioned
its reliability as an inflation hedge or source of stable real value \cite{Erb2013}. We demonstrate
that, despite these concerns, a portfolio including gold outperforms portfolios without it.

The combination of equities, fixed income, and gold also has practical precedents in investable multi-asset funds.
The permanent portfolio mutual fund (PRPFX) combines stocks, government securities, and precious metals \cite{PRPFX2026},
while the RPAR ETF allocates across equities, nominal and inflation-linked government bonds, and commodities including gold \cite{RPAR2026}.
Other multi-asset ETFs use either broad strategic allocations or momentum-based rules to vary exposure across stocks, bonds, and real assets
such as commodities, real estate, and infrastructure. Such dynamic allocations are motivated by evidence that
dynamic asset allocations can improve static ones \cite{laipply2025can}.

Scaling exposure to a target volatility is a long-standing way to improve
risk-adjusted return and attenuate the left tail of the return distribution, by
de-risking when volatility is high \cite{Moreira2017, Harvey2018,
Hocquard2013, Faber2007, Fleming2001, Kirby2012}.
The method also has industry precedents. Risk control indices scale exposure
using realized volatility and a specified volatility target
\cite{SPDJIRiskControl2026}.
This is effective because periods of high volatility often
coincide with market stress and adverse returns.
The evidence is not uniformly favorable. Across a large set of equity
strategies, volatility management does not systematically improve out-of-sample
performance \cite{Cederburg2020}.
Volatility control belongs to a broader family of risk-based allocation rules---such
as risk parity and equal risk contribution---that size positions by risk rather
than by forecast return \cite{Qian2005, Maillard2010, Asness2012}.
Risk parity and equal risk contribution portfolios choose
relative asset weights so that portfolio risk
is distributed more evenly across assets.
The same family of portfolios includes inverse volatility weighting, sometimes called
na\"{i}ve risk parity \cite{LeoteDeCarvalho2012}, risk budgeting, which splits risk
in prescribed (not necessarily equal) proportions \cite{Bruder2012, Roncalli2013},
the global minimum variance portfolio \cite{Haugen1991, Clarke2006},
the maximum diversification portfolio \cite{Choueifaty2008},
and the equal weight, or $1/N$, rule \cite{DeMiguel2009}.
Volatility control addresses a different dimension of the allocation problem.
Rather than prescribing how risk should be divided among assets,
it controls the total amount of risk taken by scaling a given portfolio
toward a specified volatility level.
We adopt volatility targeting but keep the long-only, unlevered constraints
of our setting,
so the strategy de-risks into cash in volatile regimes, but does not lever up in
calm ones.
In appendix~\ref{a-risk-based} we implement seven of these rules in our
setting, and find that none of them attains the risk-adjusted performance of
our Markowitz portfolio.

Our optimization-based portfolios combine a future return estimate with a returns covariance
estimate in the mean--variance framework of \cite{Markowitz1952}, whose canonical
scalar summary is the Sharpe ratio \cite{Sharpe1966, Sharpe1994}, and follows the
practical-implementation perspective of \cite{Black1992, Boyd2017}.
Empirical studies suggest that fixed-weight portfolios can be difficult to beat with
optimization-based portfolios, because future return forecasting is
challenging \cite{DeMiguel2009}.
In addition, Markowitz-type portfolios have received criticism for their sensitivity to
the return forecast and covariance estimate \cite{Michaud1989}.
A well-known remedy is the Black--Litterman model \cite{Black1992, He1999},
which shrinks a return forecast toward the returns implied by an equilibrium
portfolio; in appendix~\ref{a-risk-based} we compare against it, giving it our
own return forecast as its view.
We demonstrate that, with carefully constructed inputs and constraints, the optimization-based
portfolios outperform the fixed-weight benchmarks.
The return forecast used in the simple Markowitz portfolio is a simple trailing
average of past returns, which is a
common momentum signal \cite{Moskowitz2012, Hurst2017}.
Momentum signals have performed well for many decades \cite{Hurst2017}, and
scaling momentum by volatility can reduce its crash risk \cite{Barroso2015}.
The return forecast used in our Markowitz portfolio is a rolling
regression of forward returns on asset-specific and macro-financial features.
Ridge regularization stabilizes the fit when predictors are correlated or
numerous relative to the rolling sample.
Return prediction is difficult, and many macro-financial predictors perform
poorly out of sample \cite{Welch2008}.
Poor out-of-sample performance 
tends to appear more often in complicated models.
We therefore keep the model deliberately simple. 
Documented return predictors tend to decay
once published \cite{McLean2016AcademicResearch}, and complicated models are prone to over-fitting \cite{Gu2020, Bailey2017, Bailey2014DeflatedSharpe, santoni2026equity}.
More elaborate dynamic portfolio methods account directly for signal
persistence \cite{Garleanu2013}. Our method instead uses
a simpler one-period formulation.

Our optimization-based approach of first forecasting returns and then using the forecasts in order to decide on allocations
shares similarities with \cite{kim2023dynamic, mueller2024dynamic, laipply2025can}.
Kim et al.~\cite{kim2023dynamic} construct monthly U.S. growth and inflation indicators that
define economic regimes, estimate asset returns for each regime, and shift exposures toward attractive assets based on
the current regime.
Mueller-Glissmann et al.~\cite{mueller2024dynamic} use macro and market indicators to identify
regimes and then apply three overlays: moves between 60/40 and cash,
rotations between equities and bonds, and additions of commodities/gold.
Laipply et al.~\cite{laipply2025can} first estimate the correlation regime between stocks and bonds using a hidden Markov model,
and then make allocations using
mean-variance optimization based on the historic performance of the assets in the regime.

\subsection{Outline}
In~\S\ref{s-portfolios} we set our notation and describe the two fixed-weight benchmarks, their
volatility-controlled versions, and our two optimization-based portfolios.
In~\S\ref{s-results} we present the performance metrics for the six portfolios,
along with the performance of each of the assets (SPY, AGG, and GLD).
In~\S\ref{s-inflation-taxes} we evaluate the portfolios taking into account
inflation, and also taking into account a simple model of U.S.~federal taxes.
We conclude in~\S\ref{s-concl}.
In appendix~\ref{a-accounting} we give the details of the accounting we use,
and in appendix~\ref{a-alpha} we give the details of the return forecast used in
our Markowitz portfolio.
In appendix~\ref{a-significance} we analyze the statistical significance of
our results.
In appendix~\ref{a-lag} we explore the effect of lagging the data we use in
our Markowitz portfolio.
We address several questions related to our choice of the (few) hyper-parameters
in our Markowitz portfolio in appendix~\ref{a-hyper-params}.
Finally, in appendix~\ref{a-risk-based} we benchmark our Markowitz portfolio
against seven standard risk-based and Bayesian allocation methods, all of which
it outperforms, and in appendix~\ref{a-assumptions} we report the sensitivity of
our results to the assumed trading cost and cash rate.

\section{Portfolios}\label{s-portfolios}

We first discuss our notation and then introduce the six portfolios.

\subsection{Notation}\label{ss-notation}

We denote trading days (of which there are around 252 per year) by
$t = 0, 1, 2, \dots$.
We consider portfolios that
hold three assets (the ETFs SPY, AGG, and GLD), plus cash.
The (positive) value of the portfolio on trading day $t$
is denoted $V_t$, with initial value $V_0=1$.
The weights of the noncash assets on trading day $t$
are denoted $w^\mathrm{spy}_t$,
$w^\mathrm{agg}_t$, and $w^\mathrm{gld}_t$. These represent the fraction of
the total portfolio value held in each noncash asset.
They are nonnegative and satisfy
\[
w^\mathrm{spy}_t +w^\mathrm{agg}_t + w^\mathrm{gld}_t \leq 1,
\quad t=0,1,2, \ldots .
\]
The cash weight, the fraction of the portfolio value held in cash,
is
\[
c_t = 1-w^\mathrm{spy}_t -w^\mathrm{agg}_t - w^\mathrm{gld}_t,
\quad t=0,1,2, \ldots .
\]
We denote the portfolio as a weight vector
$w_t = (w^\mathrm{spy}_t,\ w^\mathrm{agg}_t,\ w^\mathrm{gld}_t)$.
Using vector notation we have, for example, $(w_t)_2 = w_t^\text{agg}$.
We can express $c_t$ as $1-\ones^T w_t$, where $\ones =(1,1,1)$.

The relative asset weights are given by
\[
\eta_t^\text{spy} = \frac{w_t^\text{spy}} {\ones^T w_t},
\qquad
\eta_t^\text{agg} = \frac{w_t^\text{agg}} {\ones^T w_t},
\qquad
\eta_t^\text{gld} = \frac{w_t^\text{gld}} {\ones^T w_t},
\]
whenever $w_t \neq 0$ (elementwise), \ie, the portfolio is not fully invested in cash.

After each day, the portfolio value and asset weights $w_t$ update according to the
returns of the assets and the prevailing risk-free interest rate on that day.
On a rebalance day, we trade to get $w_t$ to a specified target weight vector $\hat w_t$,
taking the trading cost into account.
\S\ref{ss-baselines}--\ref{ss-alphaport} describe the different methods to compute the target weight vector.
Details of the daily value and weight update are given in
appendix~\ref{a-accounting}.

\subsection{Fixed-weight benchmarks}\label{ss-baselines}

We consider two fixed-weight benchmarks:
the traditional 60/40 stock/bond portfolio, with target weight
vector $(0.6,0.4,0.0)$,
and the 50/30/20 stock/bond/gold portfolio,
with target weight vector $(0.5,0.3,0.2)$.
Following custom, these benchmark portfolios are rebalanced
to their targets annually,
on the last trading day of each calendar year.
Between rebalances the weights drift with the realized returns of the assets.
These fixed-weight benchmarks do not hold cash,
and their relative weights after each rebalance are the target portfolio weights.

\subsection{Volatility-controlled portfolios}\label{ss-vc}

From each fixed-weight benchmark we construct a volatility-controlled
portfolio, which we rebalance monthly, on the last trading day of each month.
The volatility-controlled portfolio dilutes the fixed-weight benchmark with cash to
reduce the volatility to a given target
value $\sigma^\text{tgt}$, without changing the relative asset weights compared to the fixed-weight benchmark.
When $t$ is a rebalance day, the weights are set to
\begin{equation}\label{eq-vc}
  \hat w_t = \left\{ \begin{array}{ll}
({\sigma^\mathrm{tgt}}/{\sigma_t}) w^\text{tgt},
& \sigma_t > \sigma^\text{tgt} \\
w^\text{tgt} & \sigma_t \leq \sigma^\text{tgt}.
\end{array} \right.
\end{equation}
Here $w^\text{tgt}$ is the fixed-weight benchmark portfolio (\ie,
either $(0.6,0.4,0.0)$ or $(0.5,0.3,0.2)$), and $\sigma_t$ is its estimated volatility.
The corresponding cash weight is $c_t=0$ when $\sigma_t \leq \sigma^\text{tgt}$,
and $c_t = 1-{\sigma^\mathrm{tgt}}/{\sigma_t}$
when $\sigma_t > \sigma^\text{tgt}$.
It is obvious from \eqref{eq-vc} that a volatility-controlled benchmark
preserves the relative weights at each rebalance.

We estimate the fixed-weight benchmark portfolio volatility $\sigma_t$ as the
empirical standard deviation of the benchmark's daily returns over a trailing
window of $11$ trading days, annualized by the factor $\sqrt{252}$.
A lookback period of $11$ days is short by the standards of volatility estimation,
and gives a noisy estimate of $\sigma_t$; but it is more reactive to changes in
volatility, and leads to better portfolio performance (after a modest search over various values).
Other volatility estimates, such as the Parkinson volatility estimator \cite{parkinson1980extreme} that requires low
and high portfolio values, exist
but we find that the simple trailing standard deviation works well in our setting.
These volatility-controlled portfolios are dynamic portfolios,
\ie, they change over time, because $\sigma_t$ varies with the realized returns of the assets.

\subsection{Optimization-based portfolios}\label{ss-alphaport}
\paragraph{Optimization objective.}
Our optimization-based portfolios depend on an estimate of the future asset returns,
given by the vector $\alpha_t$, and an estimate of the asset return
covariance matrix, denoted $\Sigma_t$, when $t$ corresponds to a rebalance day.
We express $\alpha_t$ as an estimate of next month's asset returns.  Let
$r^{\rm mrf}_t$ be the contemporaneous daily federal funds rate compounded over
$21$ trading days (\ie, a monthly estimate of the federal funds rate), and let $w_t$ denote the drifted risky-asset weights before
the rebalance.  For candidate risky-asset weights $w$, the corresponding cash
weight is $1-\ones^T w$, so the return on the cash is $r^{\rm mrf}_t(1-\ones^T w)$.
The estimated total gross return is then
\[
\alpha_t^T w + r^{\rm mrf}_t(1-\ones^T w).
\]

We also take the trading cost into account. We assume that the assets have the same
bid-ask spread $s$, which we take as $s=0.0005$ ($5$ basis points).
With proposed weight $w$, the estimated trading cost for trading from $w_t$ to
$w$ is then $(s/2)\|w-w_t\|_1$, where $\|x\|_1 = |x_1| + |x_2| + |x_3|$ is
the $\ell_1$ norm of a $3$-vector $x$.

Our estimate of the portfolio return over the
next month, net of the trading cost, is
\[
\alpha_t^T w + r^{\rm mrf}_t(1-\ones^T w)
-\frac{s}{2}\|w-w_t\|_1.
\]
We will maximize this estimated next-month return net of trading cost over $w$.

\paragraph{Optimization constraints.}
We start with the obvious constraints
\[
w^\mathrm{spy} \geq 0, \quad w^\mathrm{agg}\geq 0, \quad w^\mathrm{gld}
\geq 0, \qquad  w^\mathrm{spy} + w^\mathrm{agg} + w^\mathrm{gld} \le 1,
\]
which state that our asset and cash weights must be nonnegative.
These are simple linear constraints on the variable $w$.

We also impose a constraint on portfolio risk.
The estimated portfolio volatility is $(w^T \Sigma_t w)^{1/2}$. We limit
this to be below a given target volatility $\sigma^\text{tgt}$, by
imposing the constraint
\[
(w^T \Sigma_t w)^{1/2} \leq \sigma^\text{tgt}.
\]
This is a convex constraint on the variable $w$.

We also limit the relative risky-asset weight deviation from
$w^\text{tgt}=(0.5,0.3,0.2)$. Writing $\eta=w/(\ones^T w)$ when
$\ones^T w>0$, we impose
\[
\|\eta-w^\text{tgt}\|_1 = | \eta^\text{spy} - 0.5| + |\eta^\text{agg}-0.3|
+|\eta^\text{gld}-0.2| \leq 1,
\]
or, equivalently,
\BEQ\label{e-ell_1-constr}
\|w-(\ones^T w)w^\text{tgt}\|_1\leq \ones^T w.
\EEQ
This is a convex constraint on $w$.

\paragraph{Optimization problem.}
On each rebalance day $t$, we find our target portfolio weights
$\hat w_t$ by solving the optimization problem
\begin{equation}\label{eq-alphaopt}
\begin{array}{ll}
\mbox{maximize}   & \alpha_t^T w + r^{\rm mrf}_t(1-\ones^T w)
- (s/2)\|w-w_t\|_1 \\
\mbox{subject to} & w^\mathrm{spy} \geq 0, ~w^\mathrm{agg}\geq 0,~w^\mathrm{gld}
\geq 0\\ &  w^\mathrm{spy} + w^\mathrm{agg} + w^\mathrm{gld} \le 1,\\
&\left\| w - (\ones^T w) w^\text{tgt} \right\|_1 \leq \ones^T w,\\
&(w^T\Sigma_t w)^{1/2} \leq \sigma^\text{tgt},
\end{array}
\end{equation}
with variable $w \in \reals^3$.
Here we choose the weights to maximize the estimated next-month return net of
the estimated cost of the rebalance
while limiting the estimated risk to not exceed the target value and keeping
the relative risky-asset allocation within the stated $\ell_1$ distance of the
50/30/20 portfolio.
This is a small convex optimization problem \cite{Boyd2004Convex} and a variant of the classical Markowitz portfolio optimization problem \cite{Markowitz1952}.
We model the problem in CVXPY \cite{Diamond2016CVXPY}
and solve it with the Clarabel solver \cite{Goulart2026Clarabel}.
This very small problem can be solved in well under one millisecond.

\paragraph{Optimization problem data.}
The portfolio problem \eqref{eq-alphaopt}
requires the data $\alpha_t$, $r^{\rm mrf}_t$, $s$, $w_t$, $\Sigma_t$, and
$\sigma^\text{tgt}$.
The target volatility $\sigma^\text{tgt}$ is specified.
We take the bid-ask spread to be $s=0.0005$.
We take the estimated covariance $\Sigma_t$ to be the empirical covariance
matrix of the three assets' daily returns over a trailing window of
$11$ trading days, annualized.
As with our estimates of the volatilities of the fixed-weight benchmarks,
this gives a noisy but reactive estimate, which improves portfolio performance.
We use two different methods to estimate $\alpha_t$, one very simple
and one somewhat less simple.

\paragraph{Simple Markowitz.}
For the simple Markowitz portfolio, we take $\alpha_t$ to be the exponentially
weighted moving average of each asset's past daily returns, with a half-life of $252$
trading days (one year), multiplied by $21$ to put it on the next-month scale.
This is a simple momentum signal: it favors assets that
have performed well over the trailing year, and it uses only the return history.

\paragraph{Markowitz.}
For our Markowitz portfolio, $\alpha_t$ is a return forecast
based on previous asset returns, and in addition some
quantities that are publicly available on day $t$.
These fall into three groups.
\begin{itemize}
  \item \emph{Asset-specific features}, such as the trailing returns,
trading volumes, and volatilities.
  \item \emph{Macro-financial features}, such as yields, inflation rates,
and the VIX measure of expected U.S.~equity-market volatility.
  \item \emph{Fama-French factors} \cite{Fama1993},
such as the market excess return.
\end{itemize}

The forecast is produced by a rolling regression that links these
quantities to the assets' future returns; its construction is causal (at time $t$, the forecast
only depends on data available at time $t$) and is described in
appendix~\ref{a-alpha}.
Our estimation method is simple, transparent, and interpretable,
unlike more complex methods such as random forests or neural networks.

\subsection{Data and simulation}\label{ss-datasim}

We evaluate the six portfolios
over the 20 years from Jan.\ 1st 2006 to Jan.\ 1st 2026. Daily adjusted close prices for
SPY, AGG, and GLD were obtained from Yahoo Finance;
the federal funds rate (FRED
series \texttt{DFF}) and core CPI (\texttt{CPILFESL}) are downloaded from FRED.
Our simulation tracks daily portfolio value, and
includes a trading cost of $5$ basis points
(a $2.5$ basis point half-spread) on asset trading,
charged against the portfolio value at each rebalance.  The optimization-based
portfolios anticipate this same cost in their objective.
In appendix~\ref{a-assumptions} we report the sensitivity of our results to the
assumed trading cost and to the rate at which cash accrues.

\subsection{Metrics}\label{ss-metrics}

We report six metrics, all computed from the portfolio value time series.
\emph{Annualized return} is the compound annual growth rate (CAGR), \ie, the
constant annual rate of growth that takes the portfolio from its initial to its
final value over the simulation.
\emph{Annualized volatility} is the standard deviation of the daily returns,
annualized by the factor $\sqrt{252}$.
The \emph{Sharpe ratio} is the CAGR of the portfolio
value relative to the compounded federal funds rate, divided by the annualized
volatility.
This is a geometric, CAGR-based measure, and not the conventional Sharpe ratio,
which uses the arithmetic mean of the periodic excess returns.
The two agree closely here: over the full sample the conventional Sharpe ratios
are $1.08$ for the Markowitz portfolio (against the $1.08$ we report),
$0.92$ ($0.91$) for the simple Markowitz portfolio,
$0.83$ ($0.82$) for the 50/30/20 VC portfolio,
$0.73$ ($0.71$) for the 60/40 VC portfolio,
$0.73$ ($0.70$) for the 50/30/20, and
$0.60$ ($0.56$) for the 60/40.
The conventional measure is slightly more favorable to the higher-volatility
benchmarks, so the margins we report are, if anything, mildly conservative.
\emph{Maximum drawdown} (max DD) is the largest percentage decline of the portfolio
value from its previous running peak,
\ie, the worst peak-to-trough loss.
\emph{Mean drawdown} (mean DD) is the average percentage decline from the
running peak over time.
\emph{Turnover} is half the total absolute change in the non-cash weights,
so that selling one asset to buy another counts the exchanged value
once, averaged over trading days and annualized.

\section{Results}\label{s-results}

In all results reported below we use a common annualized volatility target of
$\sigma^\text{tgt} = 7\%$ for both volatility-controlled portfolios and the
optimization-based portfolios, so that the portfolios are compared at the same level of
targeted risk.  (The realized risks are, however, a bit different.)

\subsection{Metrics}
\begin{figure}
  \centering
  \includegraphics[width=0.95\linewidth]{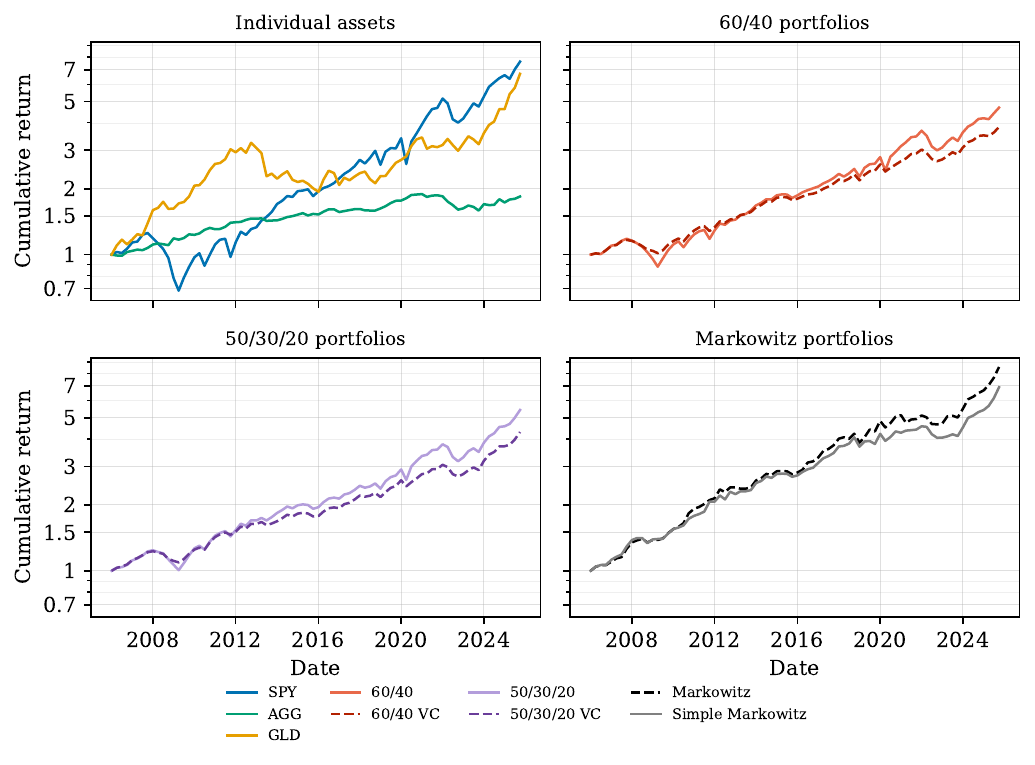}
  \caption{Cumulative return of the three assets (top left) and
the six portfolios. VC corresponds to volatility control.}
  \label{f-navs}
\end{figure}

\paragraph{Cumulative return.}
Figure~\ref{f-navs} shows the cumulative return of the six portfolios,
on a logarithmic vertical scale, so a straight line
corresponds to a constant rate of return.
The two volatility-controlled portfolios track visibly
smoother paths than their fixed-weight counterparts, particularly through the
2008 and 2020 drawdowns, and the optimization-based portfolios dominate all four
benchmarks over the full period.
Both optimization-based portfolios have a positive return in 2008, of $4\%$
for the Markowitz portfolio and $2.6\%$ for the simple Markowitz portfolio.

\paragraph{Metrics.}
Table~\ref{t-results} reports the metrics for the six portfolios.
We can see that volatility control
results in realized volatility near the target value
and reduces maximum and mean drawdown, at the cost of lower return but
higher Sharpe ratio, despite higher turnover.
The optimization-based portfolios improve the risk-adjusted returns further,
with higher Sharpe ratios and drawdowns well below the fixed-weight benchmarks.
The Markowitz portfolio achieves a Sharpe ratio of $1.08$, a substantial step up
from both the fixed-weight benchmarks and the volatility-controlled portfolios.
Accounting for realized risk, it delivers nearly twice the excess return
of the 60/40 benchmark, which has a Sharpe ratio of $0.56$.
The simple Markowitz portfolio
lands between the volatility-controlled benchmarks and the Markowitz
portfolio, with a Sharpe ratio of $0.91$.
The volatility-controlled and optimization-based portfolios have substantially higher
turnover compared to the annually rebalanced fixed-weight benchmarks;
we remind the reader that all of the results we report take into account trading costs.
The same metrics for seven standard risk-based allocation methods, run on the
same assets and data, are given in appendix~\ref{a-risk-based}; all of them fall
well short of the Markowitz portfolio.

\begin{table}
  \centering
  \small
  \setlength{\tabcolsep}{3pt}
  \begin{tabular}{lrrrrrr}
    \toprule
    Portfolio& Return & Volatility & Sharpe & Max DD & Mean DD &Turnover\\
    \midrule
    Markowitz        & 11.6\% &  9.0\% & 1.08 & 18.1\% & 3.0\% & 284.4\% \\
    Simple Markowitz & 10.4\% &  9.3\% & 0.91 & 15.8\% & 3.1\% & 278.9\% \\
    \midrule
    50/30/20 VC &  7.8\% &  7.3\% & 0.82 & 15.9\% &  2.4\% & 79.0\% \\
    60/40 VC  &  7.1\% &  7.4\% & 0.71 & 16.9\% &  2.6\% & 85.0\% \\
    \midrule
    50/30/20    &  9.1\% & 10.3\% & 0.70 & 27.1\% &   3.0\% &4.0\% \\
    60/40       &  8.1\% & 11.3\% & 0.56 & 33.7\% &   3.8\% & 3.4\% \\
    \toprule
    GLD         & 10.6\% & 17.9\% & 0.49 & 45.6\% &   17.0\% & 0.0\% \\
    AGG         &  3.1\% &  5.3\% & 0.26 & 18.4\% &   2.9\% & 0.0\% \\
    SPY         & 10.8\% & 19.4\% & 0.46 & 55.2\% &  7.5\% & 0.0\% \\
    Cash & 1.7\% & 0.0\% & 0.00 & 0.0\% &  0.0\% & 0.0\% \\
    \bottomrule
  \end{tabular}
  \caption{Performance metrics for the six portfolios (top) and
the individual assets and cash (bottom).}
  \label{t-results}
\end{table}

A detailed analysis of the statistical significance of these results
is given in appendix~\ref{a-significance}.

\begin{figure}
    \centering
        \includegraphics[width=\linewidth]{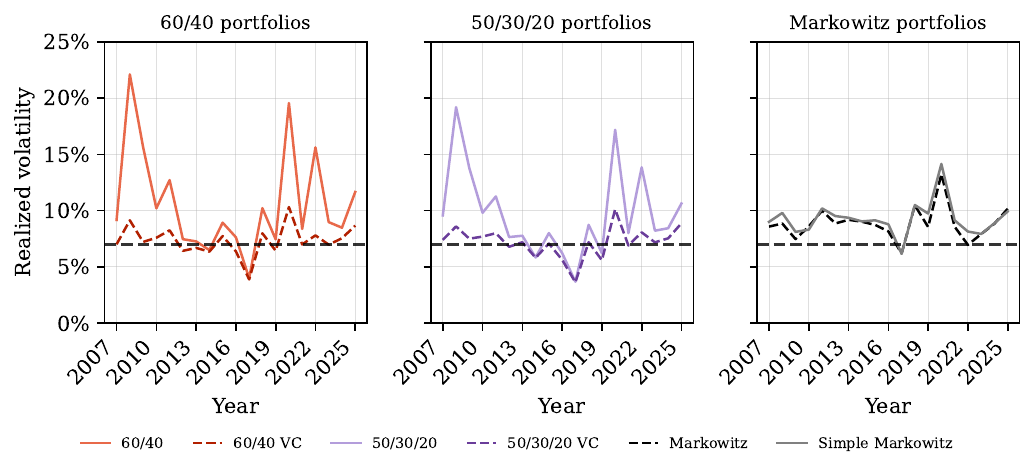}
        \caption{Calendar-year realized volatilities.}
        \label{f-yearly-volatilities}
\end{figure}

\paragraph{Consistency.}
Table~\ref{t-results} gives the aggregate performance of the portfolios over
many years.
One important additional metric not captured in these aggregate values
is the consistency of performance over time, through different
market regimes.  We illustrate this by examining the annual realized
volatilities of the six portfolios, shown
in figure~\ref{f-yearly-volatilities}.
We see that volatility control leads to much more consistent realized volatility,
as judged by the variability of realized calendar year volatility;
with the volatility-controlled and Markowitz portfolios all comparably
consistent.
The two fixed-weight benchmarks experience spikes in volatility, ranging from around 4\%
to above 20\%.
The volatility-controlled and Markowitz portfolios
bring this range down, from around 4\% to 12\%.

It is well known that return-forecasting strategies can exhibit decreasing performance
over long periods of time, presumably as other traders learn the (small) arbitrage
opportunities being exploited \cite{McLean2016AcademicResearch}.
Our Markowitz portfolio shows no clear sign of this phenomenon over our
simulation period.  Its annualized returns in the four consecutive five-year
periods are $14.0\%$, $8.0\%$, $12.5\%$, and $11.8\%$, at annualized
volatilities of $8.4\%$, $9.2\%$, $9.6\%$, and $8.6\%$, for Sharpe ratios of
$1.34$, $0.86$, $1.16$, and $0.97$; the Sharpe ratios of all six portfolios by
subperiod are given in table~\ref{t-subperiods}.

\begin{table}
  \centering
  \small
  \setlength{\tabcolsep}{3pt}
  \begin{tabular}{lrrrrr}
    \toprule
    Portfolio & 2006--2011 & 2011--2016 & 2016--2021 & 2021--2026 & 2006--2026\\
    \midrule
    Markowitz        & 1.34 & 0.86 & 1.16 & 0.97 & 1.08 \\
    Simple Markowitz & 1.11 & 0.94 & 0.82 & 0.81 & 0.91 \\
    \midrule
    50/30/20 VC      & 0.63 & 0.65 & 1.17 & 0.84 & 0.82 \\
    60/40 VC         & 0.35 & 0.99 & 0.96 & 0.57 & 0.71 \\
    \midrule
    50/30/20         & 0.40 & 0.75 & 1.05 & 0.75 & 0.70 \\
    60/40            & 0.13 & 0.99 & 0.87 & 0.48 & 0.56 \\
    \bottomrule
  \end{tabular}
  \caption{Sharpe ratios of the six portfolios by five-year subperiods, and over
the full period.}
  \label{t-subperiods}
\end{table}

\subsection{Weights over time}\label{ss-weights-time}
\begin{figure}
  \centering
  \begin{subfigure}{0.48\linewidth}
    \centering
    \includegraphics[width=\linewidth]{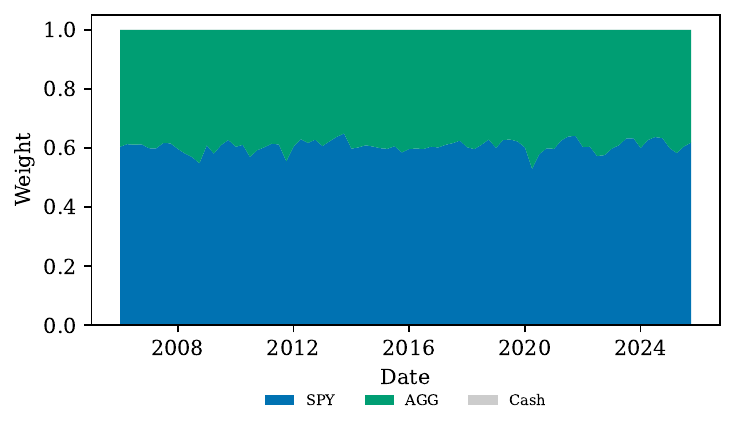}
    \caption{60/40}
    \label{f-comp-6040}
  \end{subfigure}
  \hfill
  \begin{subfigure}{0.48\linewidth}
    \centering
    \includegraphics[width=\linewidth]{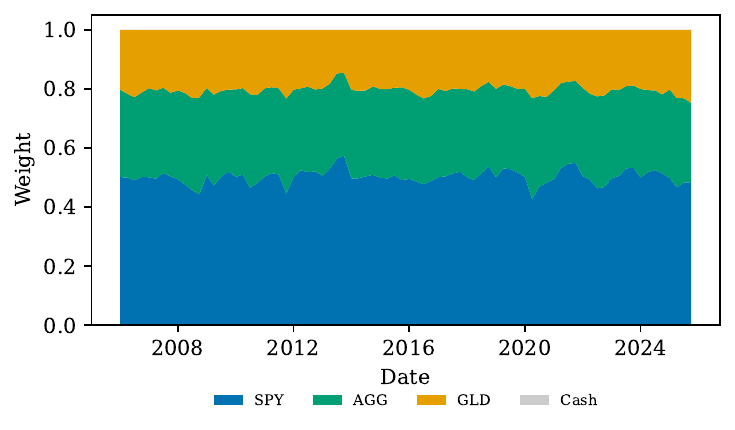}
    \caption{50/30/20}
    \label{f-comp-503020}
  \end{subfigure}

  \vspace{0.75em}

  \begin{subfigure}{0.48\linewidth}
    \centering
    \includegraphics[width=\linewidth]{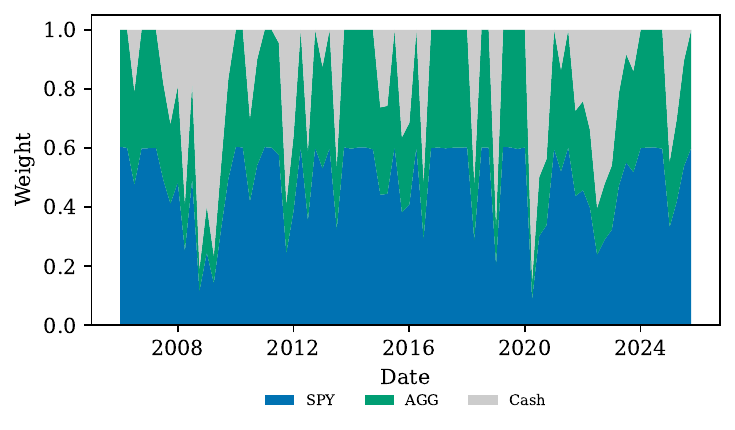}
    \caption{60/40 VC}
    \label{f-comp-6040vc}
  \end{subfigure}
  \hfill
  \begin{subfigure}{0.48\linewidth}
    \centering
    \includegraphics[width=\linewidth]{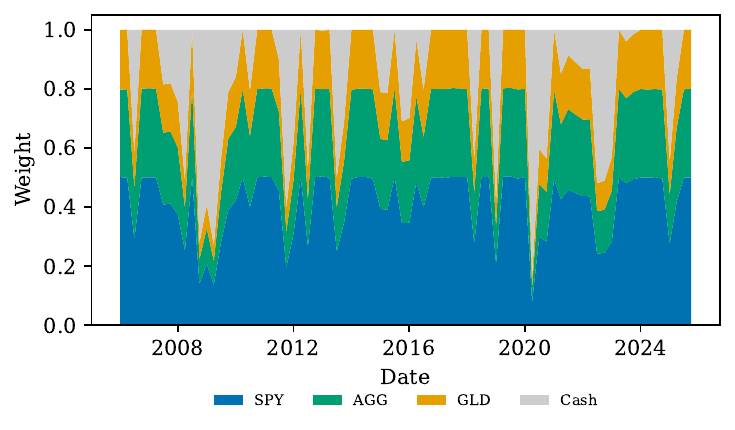}
    \caption{50/30/20 VC}
    \label{f-comp-503020vc}
  \end{subfigure}

  \vspace{0.75em}

  \begin{subfigure}{0.48\linewidth}
    \centering
    \includegraphics[width=\linewidth]{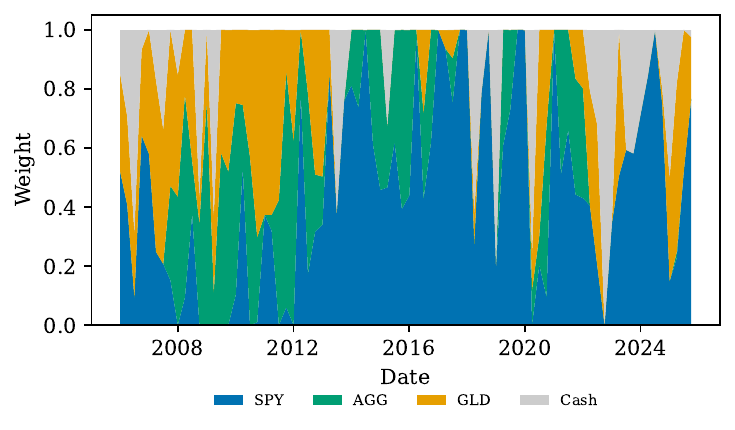}
    \caption{Simple Markowitz}
    \label{f-comp-simple-alpha}
  \end{subfigure}
  \hfill
  \begin{subfigure}{0.48\linewidth}
    \centering
    \includegraphics[width=\linewidth]{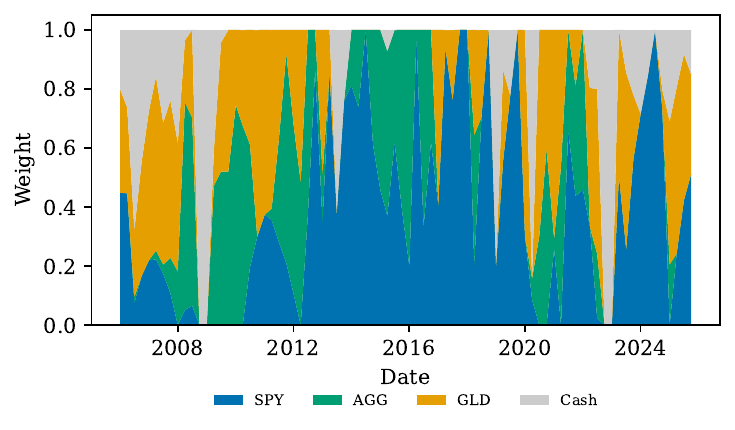}
    \caption{Markowitz}
    \label{f-comp-alpha}
  \end{subfigure}

  \caption{Portfolio weights over time.}
  \label{f-comp-all}
\end{figure}

Figure~\ref{f-comp-all} shows the weights
for the six portfolios over time. The fixed-weight
benchmarks (panels~\ref{f-comp-6040} and~\ref{f-comp-503020}) drift a bit between
annual rebalances; the volatility-controlled variants
(panels~\ref{f-comp-6040vc} and~\ref{f-comp-503020vc}) visibly scale down during
high-volatility market regimes, holding more in cash; and the Markowitz
portfolio (panel~\ref{f-comp-alpha}) shifts weight dynamically across all three
assets each month, as does the simple Markowitz portfolio
(panel~\ref{f-comp-simple-alpha}).

\subsection{Average weights over time}

Table~\ref{t-avg-weights} gives the average weights over time for each portfolio.
Interestingly, the Markowitz portfolio, which has the best performance,
holds about $15\%$ of its value in cash on average, similar to the
volatility-controlled benchmarks.
A fixed-weight portfolio with these average weights, however,
does poorly: rebalanced annually, it realizes a return of $8.5\%$ at a
volatility of $9.1\%$, for a Sharpe ratio of $0.73$---at essentially the same
volatility as the Markowitz portfolio, but well below its Sharpe ratio of
$1.08$.
The benefit of the Markowitz portfolio is therefore not simply in holding
more cash, but in varying the cash holdings (as well as the mix of stocks, bonds,
and gold) strategically.

\begin{table}
  \centering
  \small
  \setlength{\tabcolsep}{3pt}
  \begin{tabular}{lrrrr}
    \toprule
    Portfolio & SPY & AGG & GLD & Cash\\
    \midrule
    Markowitz   & 41.5\% &  20.1\% & 23.3\% & 15.1\% \\
    Simple Markowitz & 46.5\% & 19.1\% & 21.2\% & 13.2\% \\
    \midrule
    50/30/20 VC &  42.0\% &  25.1\% & 16.8\% & 16.2\% \\
    60/40 VC    &  48.9\% &  32.4\% & 0.0\% & 18.7\% \\
    \midrule
    50/30/20    &  50.3\% & 29.2\% & 20.6\% & 0.0\% \\
    60/40       &  60.8\% & 39.2\% & 0.0\% & 0.0\% \\
    \bottomrule
  \end{tabular}
  \caption{Average weights over time for the six portfolios, averaged over all
trading days in the simulation.}
  \label{t-avg-weights}
\end{table}

\subsection{Risk-return trade-off}\label{ss-risktargets}
\begin{figure}
    \centering
    \begin{subfigure}{0.49\linewidth}
        \centering
        \includegraphics[width=\linewidth]{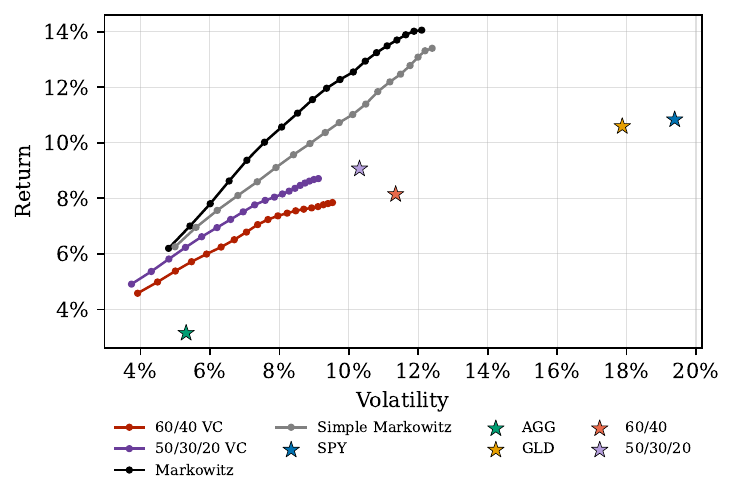}
        \caption{Realized return versus realized volatility.}
        \label{f-pareto-risk-return}
    \end{subfigure}
    \hfill
    \begin{subfigure}{0.49\linewidth}
        \centering
        \includegraphics[width=\linewidth]{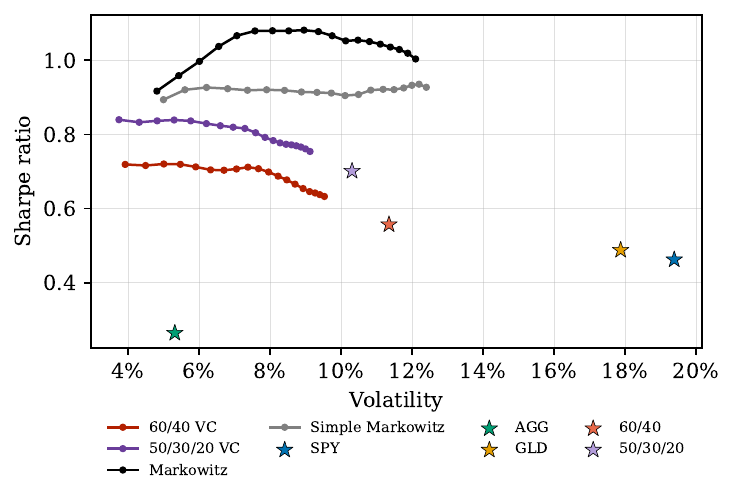}
        \caption{Sharpe ratio versus realized volatility.}
        \label{f-pareto-sharpe}
    \end{subfigure}
    \caption{
Performance of portfolios as the risk target is varied
from 3\% to 12\% in 0.5\% increments.
Raw assets and benchmarks without volatility control are shown as single stars.}
\label{f-pareto}
\end{figure}

In results reported above we target an annualized volatility of $7\%$.
Here we study the risk-return trade-off by varying the volatility target
in the volatility-controlled and optimization-based portfolios, over the range $3\%$ to
$12\%$, in $0.5\%$ increments.
Figure~\ref{f-pareto-risk-return} shows realized return versus
realized volatility for these three portfolios, and
figure~\ref{f-pareto-sharpe} shows Sharpe ratio versus realized volatility.
We can see that the Markowitz portfolio dominates the
volatility-controlled benchmarks across the entire range of volatility targets,
both in realized return at a given level of risk, and in Sharpe ratio.
The raw assets and the fixed-weight benchmarks
are shown as single points; all of them lie well below the Markowitz frontier in
both panels.

\clearpage
\section{Inflation and taxes}\label{s-inflation-taxes}

In this section we examine the post-inflation and post-tax
performance of the six portfolios.

\subsection{Inflation-adjusted performance}\label{ss-inflation}
We deflate the portfolio value by core CPI (FRED
series \texttt{CPILFESL}) and recompute the cumulative return and Sharpe ratio,
the latter now measuring return in excess of realized inflation rather than the
federal funds rate.

\begin{figure}
  \centering
  \includegraphics[width=0.95\linewidth]{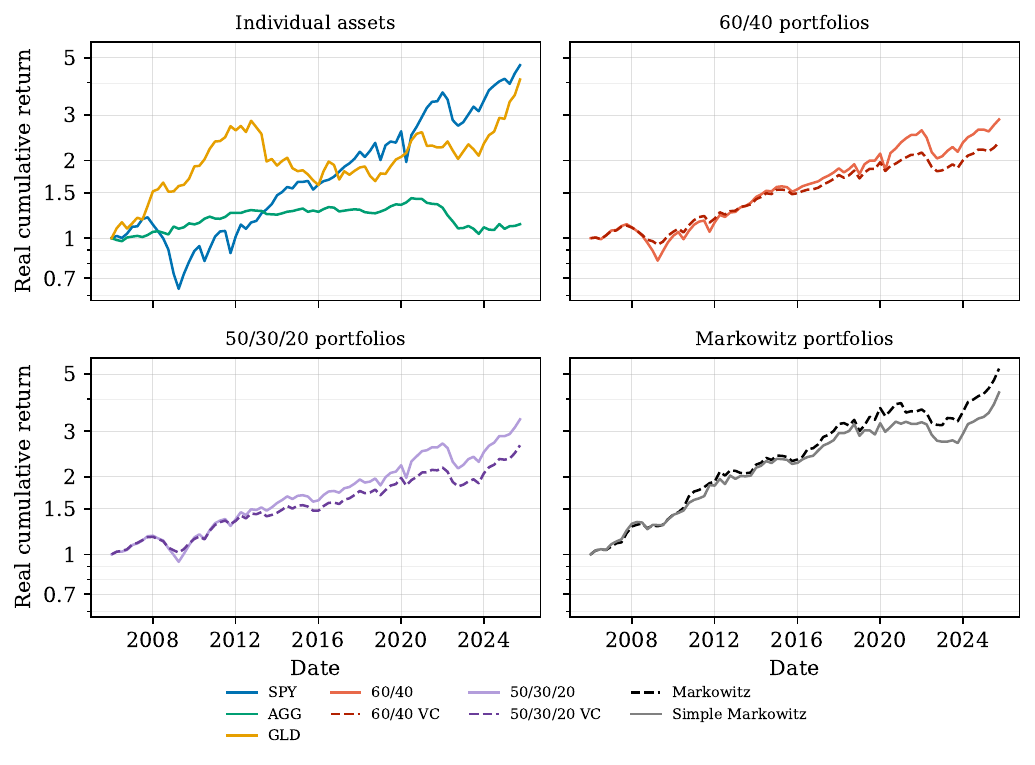}
  \caption{CPI-adjusted (real) cumulative return of the three assets (top left)
and the six portfolios.}
  \label{f-navs-real}
\end{figure}

Figure~\ref{f-navs-real} shows the CPI-adjusted (real) cumulative return of the
three assets and the six portfolios---the inflation-adjusted analog of
figure~\ref{f-navs}---and table~\ref{t-results-cpi} reports the realized annualized
return and realized Sharpe ratio. Inflation lowers every return, but the ranking is
essentially unchanged. The Markowitz portfolio retains the highest real
Sharpe ratio, $0.99$; its real return of $8.8\%$ exceeds that of raw
equity (SPY, $8.1\%$) and gold (GLD, $7.9\%$) while realizing less than half
their volatility. The advantage of the Markowitz portfolio over the
fixed-weight and volatility-controlled benchmarks thus persists after adjusting
for inflation.

\begin{table}
  \centering
  \small
  \begin{tabular}{lrr}
    \toprule
    Portfolio & Return & Sharpe \\
    \midrule
    Markowitz   & 8.8\% & 0.99 \\
    Simple Markowitz & 7.7\% & 0.82 \\
    \midrule
    50/30/20 VC & 5.1\% & 0.70 \\
    60/40 VC    & 4.4\% & 0.60 \\
    \midrule
    50/30/20    & 6.4\% & 0.62 \\
    60/40       & 5.5\% & 0.49 \\
    \midrule
    GLD         & 7.9\% & 0.44 \\
    AGG         & 0.6\% & 0.12 \\
    SPY         & 8.1\% & 0.42 \\
    \bottomrule
  \end{tabular}
  \caption{CPI-adjusted annualized return and Sharpe ratio, both in excess of
realized inflation, for the six portfolios and the three assets.}
  \label{t-results-cpi}
\end{table}

\subsection{Tax adjusted performance}\label{ss-taxes}
The results given so far are pre-tax.
In a taxable account, rebalancing realizes capital
gains (or losses), and since our dynamic portfolios trade
far more than the annually rebalanced benchmarks, we ask how much of
their advantage survives tax.
We estimate post-tax performance using a simple accounting model
described below.

\paragraph{Tax brackets.}
We consider four brackets, each with a pair of marginal rates
$(\tau^\text{lt}, \tau^\text{st})$: the long-term rate $\tau^\text{lt}$ on gains
from positions held more than one year, and the short-term (ordinary-income) rate
$\tau^\text{st}$ on positions held one year or less. They correspond to
representative U.S.~federal single-filer rates:
\begin{itemize}
\item \textbf{B1}, $(0\%,\,0\%)$: a tax-advantaged or tax-exempt account
(\eg, an IRA or 401(k)), or the lowest bracket.  This corresponds to
ignoring taxes, \ie, the pre-tax results.
\item \textbf{B2}, $(15\%,\,22\%)$: a middle-income investor.
\item \textbf{B3}, $(15\%,\,35\%)$: a high-income investor.
\item \textbf{B4}, $(20\%,\,37\%)$: the top bracket.
\end{itemize}
Two adjustments apply on top of these headline rates.
The net investment income tax (NIIT) adds a $3.8\%$ surtax on investment income
above a modified adjusted gross income of \$200,000 for a single filer, so it
applies in B3 and B4 but not in B2; the effective pairs are therefore
$(18.8\%,\,38.8\%)$ and $(23.8\%,\,40.8\%)$.
Also, GLD is a grantor trust holding physical metal, so a long-term gain on its
shares is a collectibles gain, taxed at ordinary rates capped at $28\%$ rather
than at the rate that applies to stock. Its long-term rate is thus $22\%$ in B2
and $31.8\%$ in both B3 and B4.

\paragraph{Accounting.}
We track individual tax lots. Each purchase opens a lot recording its
acquisition date and cost basis, and each sale realizes a gain or loss (proceeds
minus basis), classified as long-term or short-term by
whether the lot was held more than one year.
Whenever a transaction realizes a gain or loss at a rebalance,
we deduct the tax liability immediately from the cash account.
(In a more realistic simulation the taxes for the preceding calendar year would
be deducted from the cash account each April.)
Our results are therefore conservative: settling at once rather
than the following April gives up the use of the money in the interim, and the
more a portfolio trades the more of that deferral it forgoes.
Interest earned on cash (at the federal funds rate) is taxed as ordinary income,
\ie, at the short-term rate rather than the long-term rate.

Our simulations run on adjusted close prices, whose returns reinvest every
distribution, so a model that taxed only realized capital gains would silently
defer dividend and coupon income into capital appreciation.  We therefore
separate the income component of each asset's daily total return and tax it in
the period received: SPY's distributions as qualified dividends, at the
long-term rate, and AGG's as interest, at the ordinary rate.  (GLD makes no
distributions.)  The basis of the holding is stepped up by the amount taxed, so
the same dollars are not taxed a second time when the position is sold.
Over 2006--2025 this income averaged $1.9\%$ a year for SPY and $3.1\%$ for AGG.

We also apply the wash-sale rule.  A realized loss is deductible only if no
purchase of the same asset falls within $30$ calendar days either side of the
sale; otherwise the loss is disallowed and added to the basis of the replacement
shares.  Since consecutive monthly rebalances are $28$ to $31$ days apart, a
large share of our realized losses falls inside that window.
We ignore state and local taxes.

When we sell assets, we use a simple greedy rule to determine which lots to sell.
We choose the lots so as to minimize the immediate tax liability.
This depends on the basis of the lots, their short-term or long-term status,
and the two marginal tax rates.
We start from cash, and
at the end of the simulation horizon we
liquidate the entire portfolio and tax the remaining unrealized gains, so even a
zero-turnover holding (buy-and-hold SPY) is taxed, since its accumulated gain must
eventually be realized.

A tax-aware investor
could do better, by harvesting losses and deferring gains more aggressively
\cite{Moehle2021TaxAware}; our simple analyses here estimate the post-tax performance of a
na\"{i}ve investor who simply runs each strategy and pays the resulting
taxes.
Also, the tax analysis assumes that the investor is carrying out the rebalancing.
If the portfolios were instead offered as ETFs, the investor would only pay
taxes when shares of it are sold. This tax advantage to the investor would presumably
exceed ETF fees.

\paragraph{Results.}
Tables~\ref{t-results-tax} and~\ref{t-results-tax-sharpe} report the post-tax
annualized return and Sharpe ratio (in excess of the federal funds rate) under the
four brackets. For the Sharpe ratio we hold volatility fixed at its pre-tax (B1)
value, since the tax drag is a largely deterministic reduction in return, not
investment risk; the B1 column thus reproduces table~\ref{t-results} exactly.

\begin{table}
  \centering
  \small
  \setlength{\tabcolsep}{6pt}
  \begin{tabular}{lrrrr}
    \toprule
    Portfolio & B1 & B2 & B3 & B4 \\
    $(\tau^\text{lt}/\tau^\text{st})$ & (0/0) & (15/22) & (15/35) & (20/37) \\
    \midrule
    Markowitz   & 11.6\% &  9.2\% &  7.8\% &  7.5\% \\
    Simple Markowitz & 10.4\% & 8.4\% & 7.3\% & 7.0\% \\
    \midrule
    50/30/20 VC &  7.8\% &  6.5\% &  5.9\% &  5.6\% \\
    60/40 VC    &  7.1\% &  5.9\% &  5.5\% &  5.2\% \\
    \midrule
    50/30/20    &  9.1\% &  7.9\% &  7.4\% &  7.2\% \\
    60/40       &  8.1\% &  7.1\% &  6.7\% &  6.4\% \\
    \midrule
    GLD         & 10.6\% &  9.4\% &  8.8\% &  8.8\% \\
    AGG         &  3.1\% &  2.5\% &  1.9\% &  1.9\% \\
    SPY         & 10.8\% &  9.9\% &  9.6\% &  9.2\% \\
    \bottomrule
  \end{tabular}
  \caption{Post-tax annualized return for the six portfolios and the three
assets under the four tax brackets B1--B4, with
$(\tau^\text{lt}/\tau^\text{st})$ the long-term and short-term marginal rates in
percent.}
  \label{t-results-tax}
\end{table}

\begin{table}
  \centering
  \small
  \setlength{\tabcolsep}{6pt}
  \begin{tabular}{lrrrr}
    \toprule
    Portfolio & B1 & B2 & B3 & B4 \\
    $(\tau^\text{lt}/\tau^\text{st})$ & (0/0) & (15/22) & (15/35) & (20/37) \\
    \midrule
    Markowitz   & 1.08 & 0.83 & 0.67 & 0.64 \\
    Simple Markowitz & 0.91 & 0.71 & 0.59 & 0.56 \\
    \midrule
    50/30/20 VC & 0.82 & 0.64 & 0.56 & 0.53 \\
    60/40 VC    & 0.71 & 0.56 & 0.50 & 0.46 \\
    \midrule
    50/30/20    & 0.70 & 0.59 & 0.55 & 0.52 \\
    60/40       & 0.56 & 0.47 & 0.43 & 0.41 \\
    \midrule
    GLD         & 0.49 & 0.42 & 0.39 & 0.39 \\
    AGG         & 0.26 & 0.14 & 0.04 & 0.03 \\
    SPY         & 0.46 & 0.41 & 0.40 & 0.38 \\
    \bottomrule
  \end{tabular}
  \caption{Post-tax Sharpe ratio (in excess of the federal funds rate) for the
six portfolios and the three assets under the four tax brackets B1--B4. The
volatility in the denominator is held fixed at its pre-tax (B1) value.}
  \label{t-results-tax-sharpe}
\end{table}

Three patterns stand out. First, the tax drag grows with turnover: the annually
rebalanced benchmarks and raw assets realize gains slowly and mostly at long-term
rates, so SPY (zero turnover) loses about $1.6$ points of return at the top
bracket---most of it from the terminal liquidation---whereas the
volatility-controlled and Markowitz portfolios (turnover $79\%$--$284\%$) give up
much more.
Second, income matters as much as turnover for the bond-heavy positions. AGG
distributes about $3.1\%$ a year as interest, taxed at the ordinary rate, so its
post-tax return falls from $3.1\%$ to $1.9\%$ and its Sharpe ratio from $0.26$ to
$0.03$ at the top bracket. The zero-turnover buy-and-hold benchmark is not the
tax-efficient one; the tax-efficient asset is the one that pays no income, which
here is gold.
Third, the Markowitz portfolio's advantage narrows under taxation but remains
clear: its Sharpe ratio falls from $1.08$ to $0.83$, $0.67$, and $0.64$ across
B2--B4, yet it stays the highest in every bracket---at the top bracket it reaches
$0.64$, above the $0.56$ of the next-best portfolio. Its edge thus survives
taxation, though a high-bracket investor would weigh its higher turnover more
heavily.

\clearpage
\section{Conclusions and discussion}\label{s-concl}

\paragraph{Return forecast.}
We do not believe that our return forecasting method is the best possible;
we consider it merely good enough to provide a performance lift.
More sophisticated return forecasting could certainly be developed,
even using the same features that we use.
We also imagine that further improvement would come from using more exotic
(even if still public) data sources. However, as the complexity of the return forecast increases, so does the risk of overfitting.

\paragraph{Timing the market?}
As shown in \S\ref{ss-weights-time}, the average weights
of the Markowitz portfolio do not by themselves explain the portfolio's performance: a
fixed-weight portfolio that statically holds those average weights yields a Sharpe
ratio of only $0.73$, well below the $1.08$ of the dynamic portfolio at similar
volatility. The extra performance comes from how the portfolio varies its
exposure over time, shifting among stocks, bonds, gold, and cash from month to
month in response to the return forecast and the risk model---which is what one
would call market timing.  Consistent with this, the portfolio de-risks into cash
during turbulent periods: it is the only portfolio we consider with a positive
Sharpe ratio through the 2008 crisis, and its
realized volatility stays in a much narrower band than the fixed-weight benchmarks
(figure~\ref{f-yearly-volatilities}).

\paragraph{Comparison with other methods.}
In appendix~\ref{a-risk-based} we benchmark our portfolios against seven
standard risk-based allocation methods---among them risk parity, minimum
variance, maximum diversification, and Black--Litterman---run on the same
assets, data, costs, and constraints.
All of them land near our simple volatility-controlled benchmark, with Sharpe
ratios between $0.60$ and $0.85$, and well below the $1.08$ of the Markowitz
portfolio.
Sizing positions by risk is evidently not enough on its own; the lift comes from
combining a return forecast with a hard limit on estimated risk.

\subsection*{Acknowledgments}
We thank David Demers, Emmanuel J. Cand\`es, Trevor Hastie, Mykel J. Kochenderfer,
Logan Bell, Alexander Karolin, Oskar Wallberg,
and Daniel Cederberg for many helpful discussions.
We are especially grateful to Rishi Narang, Ronald N. Kahn, and
Maximilian Schaller
for their detailed comments and suggestions.

\clearpage
\bibliographystyle{plain}
\bibliography{refs}

@article{Brinson1986,
  author = {Brinson, Gary P. and Hood, L. Randolph and Beebower, Gilbert L.},
  title = {{Determinants of Portfolio Performance}},
  journal = {Financial Analysts Journal},
  publisher = {Informa UK Limited},
  volume = {42},
  number = {4},
  pages = {39--44},
  year = {1986},
  month = jul,
  issn = {1938-3312},
  doi = {10.2469/faj.v42.n4.39},
  url = {http://dx.doi.org/10.2469/faj.v42.n4.39},
}

@article{Brinson1991,
  author = {Brinson, Gary P. and Singer, Brian D. and Beebower, Gilbert L.},
  title = {{Determinants of Portfolio Performance II: An Update}},
  journal = {Financial Analysts Journal},
  publisher = {Informa UK Limited},
  volume = {47},
  number = {3},
  pages = {40--48},
  year = {1991},
  month = may,
  issn = {1938-3312},
  doi = {10.2469/faj.v47.n3.40},
  url = {http://dx.doi.org/10.2469/faj.v47.n3.40},
}

@article{Campbell2017,
  author = {Campbell, John Y. and Sunderam, Adi and Viceira, Luis M.},
  title = {{Inflation Bets or Deflation Hedges? The Changing Risks of Nominal Bonds}},
  journal = {Critical Finance Review},
  publisher = {Emerald},
  volume = {6},
  number = {2},
  pages = {263--301},
  year = {2017},
  month = sep,
  issn = {2164-5760},
  doi = {10.1561/104.00000043},
  url = {http://dx.doi.org/10.1561/104.00000043},
}

@article{Ilmanen2003,
  author = {Ilmanen, Antti},
  title = {{Stock-Bond Correlations}},
  journal = {The Journal of Fixed Income},
  publisher = {With Intelligence LLC},
  volume = {13},
  number = {2},
  pages = {55--66},
  year = {2003},
  month = sep,
  issn = {2168-8648},
  doi = {10.3905/jfi.2003.319353},
  url = {http://dx.doi.org/10.3905/jfi.2003.319353},
}

@article{Baur2010a,
  author = {Baur, Dirk G. and Lucey, Brian M.},
  title = {{Is Gold a Hedge or a Safe Haven? An Analysis of Stocks, Bonds and Gold}},
  journal = {Financial Review},
  publisher = {Wiley},
  volume = {45},
  number = {2},
  pages = {217--229},
  year = {2010},
  month = apr,
  issn = {1540-6288},
  doi = {10.1111/j.1540-6288.2010.00244.x},
  url = {http://dx.doi.org/10.1111/j.1540-6288.2010.00244.x},
}

@article{Baur2010b,
  author = {Baur, Dirk G. and McDermott, Thomas K.},
  title = {{Is Gold a Safe Haven? International Evidence}},
  journal = {Journal of Banking \& Finance},
  publisher = {Elsevier BV},
  volume = {34},
  number = {8},
  pages = {1886--1898},
  year = {2010},
  month = aug,
  issn = {0378-4266},
  doi = {10.1016/j.jbankfin.2009.12.008},
  url = {http://dx.doi.org/10.1016/j.jbankfin.2009.12.008},
}

@article{Faber2007,
  author = {Faber, Mebane T.},
  title = {{A Quantitative Approach to Tactical Asset Allocation}},
  journal = {The Journal of Wealth Management},
  publisher = {With Intelligence LLC},
  volume = {9},
  number = {4},
  pages = {69--79},
  year = {2007},
  month = jan,
  issn = {2374-1368},
  doi = {10.3905/jwm.2007.674809},
  url = {http://dx.doi.org/10.3905/jwm.2007.674809},
}

@article{Moreira2017,
  author = {Moreira, Alan and Muir, Tyler},
  title = {{Volatility-Managed Portfolios}},
  journal = {The Journal of Finance},
  publisher = {Wiley},
  volume = {72},
  number = {4},
  pages = {1611--1644},
  year = {2017},
  month = may,
  issn = {1540-6261},
  doi = {10.1111/jofi.12513},
  url = {http://dx.doi.org/10.1111/jofi.12513},
}

@article{Harvey2018,
  author = {Harvey, Campbell R. and Hoyle, Edward and Korgaonkar, Russell and Rattray, Sandy and Sargaison, Matthew and Van Hemert, Otto},
  title = {{The Impact of Volatility Targeting}},
  journal = {The Journal of Portfolio Management},
  publisher = {With Intelligence LLC},
  volume = {45},
  number = {1},
  pages = {14--33},
  year = {2018},
  month = oct,
  issn = {2168-8656},
  doi = {10.3905/jpm.2018.45.1.014},
  url = {http://dx.doi.org/10.3905/jpm.2018.45.1.014},
}

@article{Hocquard2013,
  author = {Hocquard, Alexandre and Ng, Sunny and Papageorgiou, Nicolas},
  title = {{A Constant-Volatility Framework for Managing Tail Risk}},
  journal = {The Journal of Portfolio Management},
  publisher = {With Intelligence LLC},
  volume = {39},
  number = {2},
  pages = {28--40},
  year = {2013},
  month = jan,
  issn = {2168-8656},
  doi = {10.3905/jpm.2013.39.2.028},
  url = {http://dx.doi.org/10.3905/jpm.2013.39.2.028},
}

@techreport{Qian2005,
  author = {Qian, Edward E.},
  title = {{Risk Parity Portfolios: Efficient Portfolios Through True Diversification}},
  institution = {PanAgora Asset Management},
  type = {White Paper},
  year = {2005},
  month = sep,
  url = {https://www.panagora.com/assets/PanAgora-Risk-Parity-Portfolios-Efficient-Portfolios-Through-True-Diversification.pdf},
}

@article{Maillard2010,
  author = {Maillard, S\'ebastien and Roncalli, Thierry and Te\"iletche, J\'er\^ome},
  title = {{The Properties of Equally Weighted Risk Contribution Portfolios}},
  journal = {The Journal of Portfolio Management},
  publisher = {With Intelligence LLC},
  volume = {36},
  number = {4},
  pages = {60--70},
  year = {2010},
  month = jul,
  issn = {2168-8656},
  doi = {10.3905/jpm.2010.36.4.060},
  url = {http://dx.doi.org/10.3905/jpm.2010.36.4.060},
}

@article{Asness2012,
  author = {Asness, Clifford S. and Frazzini, Andrea and Pedersen, Lasse H.},
  title = {{Leverage Aversion and Risk Parity}},
  journal = {Financial Analysts Journal},
  publisher = {Informa UK Limited},
  volume = {68},
  number = {1},
  pages = {47--59},
  year = {2012},
  month = jan,
  issn = {1938-3312},
  doi = {10.2469/faj.v68.n1.1},
  url = {http://dx.doi.org/10.2469/faj.v68.n1.1},
}

@article{Fama1993,
  author = {Fama, Eugene F. and French, Kenneth R.},
  title = {{Common Risk Factors in the Returns on Stocks and Bonds}},
  journal = {Journal of Financial Economics},
  publisher = {Elsevier BV},
  volume = {33},
  number = {1},
  pages = {3--56},
  year = {1993},
  month = feb,
  issn = {0304-405X},
  doi = {10.1016/0304-405X(93)90023-5},
  url = {http://dx.doi.org/10.1016/0304-405X(93)90023-5},
}

@article{Gu2020,
  author = {Gu, Shihao and Kelly, Bryan and Xiu, Dacheng},
  editor = {Karolyi, Andrew},
  title = {{Empirical Asset Pricing via Machine Learning}},
  journal = {The Review of Financial Studies},
  publisher = {Oxford University Press (OUP)},
  volume = {33},
  number = {5},
  pages = {2223--2273},
  year = {2020},
  month = feb,
  issn = {1465-7368},
  doi = {10.1093/rfs/hhaa009},
  url = {http://dx.doi.org/10.1093/rfs/hhaa009},
}

@article{Markowitz1952,
  author = {Markowitz, Harry},
  title = {{Portfolio Selection}},
  journal = {The Journal of Finance},
  publisher = {Wiley},
  volume = {7},
  number = {1},
  pages = {77--91},
  year = {1952},
  month = mar,
  issn = {1540-6261},
  doi = {10.1111/j.1540-6261.1952.tb01525.x},
  url = {http://dx.doi.org/10.1111/j.1540-6261.1952.tb01525.x},
}

@article{Black1992,
  author = {Black, Fischer and Litterman, Robert},
  title = {{Global Portfolio Optimization}},
  journal = {Financial Analysts Journal},
  publisher = {Informa UK Limited},
  volume = {48},
  number = {5},
  pages = {28--43},
  year = {1992},
  month = sep,
  issn = {1938-3312},
  doi = {10.2469/faj.v48.n5.28},
  url = {http://dx.doi.org/10.2469/faj.v48.n5.28},
}

@article{Boyd2017,
  author = {Boyd, Stephen and Diamond, Steven and Koh, Kwangmoo and Nystrup, Peter and Busseti, Enzo and Kahn, Ronald N. and Speth, Jan},
  title = {{Multi-Period Trading via Convex Optimization}},
  journal = {Foundations and Trends in Optimization},
  publisher = {Emerald},
  volume = {3},
  number = {1},
  pages = {1--76},
  year = {2017},
  month = aug,
  issn = {2167-3918},
  doi = {10.1561/2400000023},
  url = {http://dx.doi.org/10.1561/2400000023},
}

@article{Sharpe1966,
  author = {Sharpe, William F.},
  title = {{Mutual Fund Performance}},
  journal = {The Journal of Business},
  publisher = {University of Chicago Press},
  volume = {39},
  number = {1},
  pages = {119--138},
  year = {1966},
  issn = {00219398, 15375374},
  url = {http://www.jstor.org/stable/2351741},
}

@article{Sharpe1994,
  author = {Sharpe, William F.},
  title = {{The Sharpe Ratio}},
  journal = {The Journal of Portfolio Management},
  publisher = {With Intelligence LLC},
  volume = {21},
  number = {1},
  pages = {49--58},
  year = {1994},
  month = oct,
  issn = {2168-8656},
  doi = {10.3905/jpm.1994.409501},
  url = {http://dx.doi.org/10.3905/jpm.1994.409501},
}

@book{Bernstein2000,
  author = {Bernstein, William J.},
  title = {{The Intelligent Asset Allocator: How to Build Your Portfolio to Maximize Returns and Minimize Risk}},
  publisher = {McGraw-Hill Education},
  year = {2000},
  isbn = {9780071362368},
  url = {https://books.google.com/books?id=ZpVrUWhn7sMC},
}

@book{Bogle2017,
  author = {Bogle, John C.},
  title = {{The Little Book of Common Sense Investing: The Only Way to Guarantee Your Fair Share of Stock Market Returns}},
  series = {Little Books. Big Profits},
  publisher = {Wiley},
  year = {2017},
  isbn = {9781119404507},
  url = {https://books.google.com/books?id=Vrg1DwAAQBAJ},
}

@techreport{Fink2025,
  author = {Fink, Laurence D.},
  title = {{2025 Annual Chairman's Letter to Investors}},
  institution = {BlackRock, Inc.},
  type = {Annual Letter},
  year = {2025},
  month = apr,
  url = {https://www.blackrock.com/corporate/investor-relations/2025-larry-fink-annual-chairmans-letter},
}

@misc{Yasmin2025,
  author = {Yasmin, Mehnaz},
  title = {{Morgan Stanley CIO Favors 60/20/20 Portfolio Strategy with Gold Inflation Hedge}},
  howpublished = {Reuters Markets},
  year = {2025},
  month = sep,
  day = {16},
  url = {https://www.reuters.com/markets/wealth/morgan-stanley-cio-favors-602020-portfolio-strategy-with-gold-inflation-hedge-2025-09-16/},
}

@article{Brixton2023,
  author = {Brixton, Alfie and Brooks, Jordan and Hecht, Pete and Ilmanen, Antti and Maloney, Thomas and McQuinn, Nicholas},
  title = {{A Changing Stock--Bond Correlation: Drivers and Implications}},
  journal = {The Journal of Portfolio Management},
  publisher = {With Intelligence LLC},
  volume = {49},
  number = {4},
  pages = {64--80},
  year = {2023},
  month = jan,
  issn = {2168-8656},
  doi = {10.3905/jpm.2023.1.459},
  url = {http://dx.doi.org/10.3905/jpm.2023.1.459},
}

@article{Boyd2024,
  author = {Boyd, Stephen and Johansson, Kasper and Kahn, Ronald and Schiele, Philipp and Schmelzer, Thomas},
  title = {{Markowitz Portfolio Construction at Seventy}},
  journal = {The Journal of Portfolio Management},
  publisher = {With Intelligence LLC},
  volume = {50},
  number = {8},
  pages = {117--160},
  year = {2024},
  month = jun,
  issn = {2168-8656},
  doi = {10.3905/jpm.2024.50.8.117},
  url = {http://dx.doi.org/10.3905/jpm.2024.50.8.117},
}

@book{Boyd2004Convex,
  author = {Boyd, Stephen and Vandenberghe, Lieven},
  title = {{Convex Optimization}},
  publisher = {Cambridge University Press},
  year = {2004},
}

@article{Goulart2026Clarabel,
  author = {Goulart, Paul J. and Chen, Yuwen},
  title = {{Clarabel: An Interior-Point Solver for Conic Programs with Quadratic Objectives}},
  journal = {Mathematical Programming Computation},
  publisher = {Springer Science and Business Media LLC},
  year = {2026},
  month = may,
  issn = {1867-2957},
  doi = {10.1007/s12532-026-00320-7},
  url = {http://dx.doi.org/10.1007/s12532-026-00320-7},
}

@article{Diamond2016CVXPY,
  author = {Diamond, Steven and Boyd, Stephen},
  title = {{CVXPY: A Python-Embedded Modeling Language for Convex Optimization}},
  journal = {Journal of Machine Learning Research},
  volume = {17},
  number = {83},
  pages = {1--5},
  year = {2016},
  url = {http://jmlr.org/papers/v17/15-408.html},
}

@misc{Smith2025LPL6040,
  author = {Smith, George},
  title = {{Is the 60/40 Portfolio Still Relevant? Exploring Alternatives}},
  howpublished = {LPL Research Blog},
  year = {2025},
  month = dec,
  day = {11},
  url = {https://www.lpl.com/research/blog/is-the-60-40-portfolio-still-relevant-exploring-alternatives.html},
  note = {Additional content provided by Kent Cullinane, CFA. Accessed: 2026-06-07},
}

@misc{Devanathan2026SingleAsset,
  author = {Devanathan, Nikhil and Rueter, Dylan and Boyd, Stephen and Cand\`es, Emmanuel and Hastie, Trevor and Kochenderfer, Mykel J. and Apoorv, Arpit and Soronow, David and Zamkovsky, Igor},
  title = {{Single-Asset Adaptive Leveraged Volatility Control}},
  howpublished = {arXiv:2603.01298v2 [q-fin.PM]},
  publisher = {arXiv},
  year = {2026},
  month = mar,
  doi = {10.48550/ARXIV.2603.01298},
  url = {https://arxiv.org/abs/2603.01298v2},
}

@article{Moehle2021TaxAware,
  author = {Moehle, Nicholas and Kochenderfer, Mykel J. and Boyd, Stephen and Ang, Andrew},
  title = {{Tax-Aware Portfolio Construction via Convex Optimization}},
  journal = {Journal of Optimization Theory and Applications},
  publisher = {Springer Science and Business Media LLC},
  volume = {189},
  number = {2},
  pages = {364--383},
  year = {2021},
  month = may,
  issn = {1573-2878},
  doi = {10.1007/s10957-021-01823-0},
  url = {http://dx.doi.org/10.1007/s10957-021-01823-0},
}

@article{McLean2016AcademicResearch,
  author = {McLean, R. David and Pontiff, Jeffrey},
  title = {{Does Academic Research Destroy Stock Return Predictability?}},
  journal = {The Journal of Finance},
  publisher = {Wiley},
  volume = {71},
  number = {1},
  pages = {5--32},
  year = {2016},
  month = jan,
  issn = {1540-6261},
  doi = {10.1111/jofi.12365},
  url = {http://dx.doi.org/10.1111/jofi.12365},
}

@article{DeMiguel2009,
  author = {DeMiguel, Victor and Garlappi, Lorenzo and Uppal, Raman},
  title = {{Optimal Versus Naive Diversification: How Inefficient Is the 1/N Portfolio Strategy?}},
  journal = {The Review of Financial Studies},
  publisher = {Oxford University Press (OUP)},
  volume = {22},
  number = {5},
  pages = {1915--1953},
  year = {2009},
  month = may,
  issn = {1465-7368},
  doi = {10.1093/rfs/hhm075},
  url = {http://dx.doi.org/10.1093/rfs/hhm075},
}

@article{Moskowitz2012,
  author = {Moskowitz, Tobias J. and Ooi, Yao Hua and Pedersen, Lasse Heje},
  title = {{Time Series Momentum}},
  journal = {Journal of Financial Economics},
  publisher = {Elsevier BV},
  volume = {104},
  number = {2},
  pages = {228--250},
  year = {2012},
  month = may,
  issn = {0304-405X},
  doi = {10.1016/j.jfineco.2011.11.003},
  url = {http://dx.doi.org/10.1016/j.jfineco.2011.11.003},
}

@article{Hurst2017,
  author = {Hurst, Brian and Ooi, Yao Hua and Pedersen, Lasse Heje},
  title = {{A Century of Evidence on Trend-Following Investing}},
  journal = {The Journal of Portfolio Management},
  publisher = {With Intelligence LLC},
  volume = {44},
  number = {1},
  pages = {15--29},
  year = {2017},
  month = oct,
  issn = {2168-8656},
  doi = {10.3905/jpm.2017.44.1.015},
  url = {http://dx.doi.org/10.3905/jpm.2017.44.1.015},
}

@article{Erb2013,
  author = {Erb, Claude B. and Harvey, Campbell R.},
  title = {{The Golden Dilemma}},
  journal = {Financial Analysts Journal},
  publisher = {Informa UK Limited},
  volume = {69},
  number = {4},
  pages = {10--42},
  year = {2013},
  month = jul,
  issn = {1938-3312},
  doi = {10.2469/faj.v69.n4.1},
  url = {http://dx.doi.org/10.2469/faj.v69.n4.1},
}

@article{Michaud1989,
  author = {Michaud, Richard O.},
  title = {{The Markowitz Optimization Enigma: Is ``Optimized'' Optimal?}},
  journal = {Financial Analysts Journal},
  publisher = {Informa UK Limited},
  volume = {45},
  number = {1},
  pages = {31--42},
  year = {1989},
  month = jan,
  issn = {1938-3312},
  doi = {10.2469/faj.v45.n1.31},
  url = {http://dx.doi.org/10.2469/faj.v45.n1.31},
}

@article{Bailey2017,
  author = {Bailey, David H. and Borwein, Jonathan M. and L\'opez de Prado, Marcos and Zhu, Qiji Jim},
  title = {{The Probability of Backtest Overfitting}},
  journal = {The Journal of Computational Finance},
  publisher = {Infopro Digital Services Limited},
  volume = {20},
  number = {4},
  pages = {39--69},
  year = {2017},
  month = apr,
  issn = {1460-1559},
  doi = {10.21314/jcf.2016.322},
  url = {http://dx.doi.org/10.21314/jcf.2016.322},
}

@article{Bailey2014DeflatedSharpe,
  author = {Bailey, David H. and L\'opez de Prado, Marcos},
  title = {{The Deflated Sharpe Ratio: Correcting for Selection Bias, Backtest Overfitting, and Non-Normality}},
  journal = {The Journal of Portfolio Management},
  publisher = {With Intelligence LLC},
  volume = {40},
  number = {5},
  pages = {94--107},
  year = {2014},
  month = sep,
  issn = {2168-8656},
  doi = {10.3905/jpm.2014.40.5.094},
  url = {http://dx.doi.org/10.3905/jpm.2014.40.5.094},
}

@article{mueller2024dynamic,
  title={Dynamic {A}sset {A}llocation using {M}achine {L}earning: {S}eeing the {F}orest for the {T}rees},
  author={Mueller-Glissmann, Christian and Ferrario, Andrea},
  journal={Journal of Portfolio Management},
  volume={50},
  number={5},
  pages={132},
  year={2024}
}

@article{kim2023dynamic,
  title={Dynamic {A}sset {A}llocation {S}trategy: {A}n {E}conomic {R}egime {A}pproach},
  author={Kim, Min Jeong and Kwon, Dohyoung},
  journal={Journal of Asset Management},
  volume={24},
  number={2},
  pages={136--147},
  year={2023},
  publisher={Springer}
}

@article{parkinson1980extreme,
  title={The {E}xtreme {V}alue {M}ethod for {E}stimating the {V}ariance of the {R}ate of {R}eturn},
  author={Parkinson, Michael},
  journal={Journal of business},
  pages={61--65},
  year={1980},
  publisher={JSTOR}
}

@article{Choueifaty2008,
  author = {Choueifaty, Yves and Coignard, Yves},
  title = {{Toward Maximum Diversification}},
  journal = {The Journal of Portfolio Management},
  publisher = {With Intelligence LLC},
  volume = {35},
  number = {1},
  pages = {40--51},
  year = {2008},
  month = oct,
  issn = {2168-8656},
  doi = {10.3905/jpm.2008.35.1.40},
  url = {http://dx.doi.org/10.3905/jpm.2008.35.1.40},
}

@article{Clarke2006,
  author = {Clarke, Roger G. and de Silva, Harindra and Thorley, Steven},
  title = {{Minimum-Variance Portfolios in the U.S. Equity Market}},
  journal = {The Journal of Portfolio Management},
  publisher = {With Intelligence LLC},
  volume = {33},
  number = {1},
  pages = {10--24},
  year = {2006},
  month = oct,
  issn = {2168-8656},
  doi = {10.3905/jpm.2006.661366},
  url = {http://dx.doi.org/10.3905/jpm.2006.661366},
}

@article{Haugen1991,
  author = {Haugen, Robert A. and Baker, Nardin L.},
  title = {{The Efficient Market Inefficiency of Capitalization-Weighted Stock Portfolios}},
  journal = {The Journal of Portfolio Management},
  publisher = {With Intelligence LLC},
  volume = {17},
  number = {3},
  pages = {35--40},
  year = {1991},
  month = apr,
  issn = {2168-8656},
  doi = {10.3905/jpm.1991.409335},
  url = {http://dx.doi.org/10.3905/jpm.1991.409335},
}

@article{LeoteDeCarvalho2012,
  author = {Leote de Carvalho, Raul and Lu, Xiao and Moulin, Pierre},
  title = {{Demystifying Equity Risk-Based Strategies: A Simple Alpha Plus Beta Description}},
  journal = {The Journal of Portfolio Management},
  publisher = {With Intelligence LLC},
  volume = {38},
  number = {3},
  pages = {56--70},
  year = {2012},
  month = apr,
  issn = {2168-8656},
  doi = {10.3905/jpm.2012.38.3.056},
  url = {http://dx.doi.org/10.3905/jpm.2012.38.3.056},
}

@techreport{Bruder2012,
  author = {Bruder, Benjamin and Roncalli, Thierry},
  title = {{Managing Risk Exposures Using the Risk Budgeting Approach}},
  institution = {Lyxor Asset Management},
  type = {Working Paper},
  year = {2012},
  doi = {10.2139/ssrn.2009778},
  url = {http://dx.doi.org/10.2139/ssrn.2009778},
}

@book{Roncalli2013,
  author = {Roncalli, Thierry},
  title = {{Introduction to Risk Parity and Budgeting}},
  publisher = {Chapman and Hall/CRC},
  series = {Chapman and Hall/CRC Financial Mathematics Series},
  year = {2013},
  doi = {10.1201/b15151},
  url = {http://dx.doi.org/10.1201/b15151},
}

@techreport{He1999,
  author = {He, Guangliang and Litterman, Robert},
  title = {{The Intuition Behind Black-Litterman Model Portfolios}},
  institution = {Goldman Sachs Investment Management Research},
  type = {Working Paper},
  year = {1999},
  doi = {10.2139/ssrn.334304},
  url = {http://dx.doi.org/10.2139/ssrn.334304},
}

@article{Fleming2001,
  author = {Fleming, Jeff and Kirby, Chris and Ostdiek, Barbara},
  title = {{The Economic Value of Volatility Timing}},
  journal = {The Journal of Finance},
  publisher = {Wiley},
  volume = {56},
  number = {1},
  pages = {329--352},
  year = {2001},
  month = feb,
  issn = {1540-6261},
  doi = {10.1111/0022-1082.00327},
  url = {http://dx.doi.org/10.1111/0022-1082.00327},
}

@article{Kirby2012,
  author = {Kirby, Chris and Ostdiek, Barbara},
  title = {{It's All in the Timing: Simple Active Portfolio Strategies that Outperform Na\"ive Diversification}},
  journal = {Journal of Financial and Quantitative Analysis},
  publisher = {Cambridge University Press},
  volume = {47},
  number = {2},
  pages = {437--467},
  year = {2012},
  month = apr,
  issn = {1756-6916},
  doi = {10.1017/S0022109012000117},
  url = {http://dx.doi.org/10.1017/S0022109012000117},
}

@article{Cederburg2020,
  author = {Cederburg, Scott and O'Doherty, Michael S. and Wang, Feifei and Yan, Xuemin (Sterling)},
  title = {{On the Performance of Volatility-Managed Portfolios}},
  journal = {Journal of Financial Economics},
  publisher = {Elsevier BV},
  volume = {138},
  number = {1},
  pages = {95--117},
  year = {2020},
  month = oct,
  issn = {0304-405X},
  doi = {10.1016/j.jfineco.2020.04.015},
  url = {http://dx.doi.org/10.1016/j.jfineco.2020.04.015},
}

@article{Politis1994,
  author = {Politis, Dimitris N. and Romano, Joseph P.},
  title = {{The Stationary Bootstrap}},
  journal = {Journal of the American Statistical Association},
  publisher = {Informa UK Limited},
  volume = {89},
  number = {428},
  pages = {1303--1313},
  year = {1994},
  month = dec,
  issn = {1537-274X},
  doi = {10.1080/01621459.1994.10476870},
  url = {http://dx.doi.org/10.1080/01621459.1994.10476870},
}

@article{Barroso2015,
  author = {Barroso, Pedro and Santa-Clara, Pedro},
  title = {{Momentum Has Its Moments}},
  journal = {Journal of Financial Economics},
  publisher = {Elsevier BV},
  volume = {116},
  number = {1},
  pages = {111--120},
  year = {2015},
  month = apr,
  issn = {0304-405X},
  doi = {10.1016/j.jfineco.2014.11.010},
  url = {http://dx.doi.org/10.1016/j.jfineco.2014.11.010},
}

@article{Welch2008,
  author = {Welch, Ivo and Goyal, Amit},
  title = {{A Comprehensive Look at The Empirical Performance of Equity Premium Prediction}},
  journal = {The Review of Financial Studies},
  publisher = {Oxford University Press (OUP)},
  volume = {21},
  number = {4},
  pages = {1455--1508},
  year = {2008},
  month = jul,
  issn = {1465-7368},
  doi = {10.1093/rfs/hhm014},
  url = {http://dx.doi.org/10.1093/rfs/hhm014},
}

@article{Garleanu2013,
  author = {G{\^a}rleanu, Nicolae and Pedersen, Lasse Heje},
  title = {{Dynamic Trading with Predictable Returns and Transaction Costs}},
  journal = {The Journal of Finance},
  publisher = {Wiley},
  volume = {68},
  number = {6},
  pages = {2309--2340},
  year = {2013},
  month = dec,
  issn = {1540-6261},
  doi = {10.1111/jofi.12080},
  url = {http://dx.doi.org/10.1111/jofi.12080},
}

@article{Hillier2006,
  author = {Hillier, David and Draper, Paul and Faff, Robert},
  title = {{Do Precious Metals Shine? An Investment Perspective}},
  journal = {Financial Analysts Journal},
  publisher = {Informa UK Limited},
  volume = {62},
  number = {2},
  pages = {98--106},
  year = {2006},
  month = mar,
  issn = {1938-3312},
  doi = {10.2469/faj.v62.n2.4085},
  url = {http://dx.doi.org/10.2469/faj.v62.n2.4085},
}

@misc{PRPFX2026,
  author = {{Permanent Portfolio Family of Funds}},
  title = {{Permanent Portfolio}},
  year = {2026},
  url = {https://www.prpfx.com/permanent-portfolio.html},
  note = {Accessed: 2026-09-01},
}

@misc{RPAR2026,
  author = {{RPAR Risk Parity ETF}},
  title = {{RPAR Risk Parity ETF}},
  year = {2026},
  url = {https://www.rparetf.com/rpar},
  note = {Accessed: 2026-09-01},
}

@misc{SPDJIRiskControl2026,
  author = {{S\&P Dow Jones Indices}},
  title = {{Risk Control Indices}},
  year = {2026},
  url = {https://www.spglobal.com/spdji/en/index-family/commodities/quantitative-strategies/risk-control/},
  note = {Accessed: 2026-09-01},
}

@article{laipply2025can,
  title={Can {B}onds {S}till {D}iversify {M}ulti-{A}sset {P}ortfolios? {I}ncome {V}ersus {D}uration in {D}istinct {C}orrelation {R}egimes},
  author={Laipply, Steve and Madhavan, Ananth},
  journal={Income versus Duration in Distinct Correlation Regimes (September 15, 2025)},
  year={2025}
}

@article{santoni2026equity,
  title={Equity {S}trategy {B}acktesting: {L}uck or {E}dge? The {M}inervaScore as a {S}tatistical {R}obustness {G}rade},
  author={Santoni, Maria Laura and Jouanne, Vincent and Scullin, Matthew L},
  journal={arXiv preprint arXiv:2608.23808},
  year={2026}
}

\clearpage
\appendix

\section{Portfolio accounting}\label{a-accounting}

Let $r_t \in \reals^3$ be the vector of close-to-close daily simple returns on day
$t$, and let $r^\text{rf}_t$ be the daily risk-free rate
(the federal funds rate expressed per trading day). The cash weight is
$c_t = 1 - (w^\mathrm{spy}_t + w^\mathrm{agg}_t + w^\mathrm{gld}_t)$.

On a day that is \emph{not} a rebalance day, each asset holding grows with its
return and cash grows at the risk-free rate. The portfolio value $V_t$ updates as
\[
V_t = V_{t-1}\left(\sum_{i=1}^3 w^i_{t-1}\,(1 + r^i_t) +
c_{t-1}\,(1 + r^\text{rf}_t)\right),
\]
and the drifted weights become
\[
(w_t)_i = \frac{V_{t-1}}{V_t}\, w^i_{t-1}\,(1 + r^i_t),
\quad i =1,2,3.
\]
The numerator is the value of the $i$th asset after its return, and the denominator is
the new portfolio value, so the ratio is the new weight of the $i$th asset.

On a rebalance day, the value and weights are first updated as above, giving
the drifted weights $w_t$ and the pre-trade value $V_t$. We then compute the target weights $\hat w_t$
by one of the
methods in \S\ref{s-portfolios}. We finally trade so
that the weights equal the desired weights $\hat w_t$; 
so the post-trade cash weight is
$\hat c_t = 1 - (\hat w^\mathrm{spy}_t + \hat w^\mathrm{agg}_t
+ \hat w^\mathrm{gld}_t)$.

The total traded fraction of the portfolio is
$\sum_{i=1}^3 |\hat w^i_t - w^i_t|$, counting the three (noncash) assets
only; moving value into or out of cash is not itself a trade. Trading incurs
a trading cost at the half-spread $s/2$. With $s = 5$ basis points, the
trading cost of a rebalance is
\[
\frac{s}{2} \sum_{i=1}^3 \left| \hat w^i_t - w^i_t \right| V_t .
\]
This cost is deducted from the cash account, so the post-trade value is the
pre-trade value minus the trading cost, and the post-trade weights are
re-computed using the updated portfolio value.

We define the turnover of a rebalance as half the total traded fraction,
$\tfrac{1}{2} \sum_{i=1}^3 |\hat w^i_t - w^i_t|$, so that selling one asset
to buy another counts the exchanged value once.
We report annualized turnover as $252$ times the average daily turnover,
with zero turnover on non-rebalancing days.

\clearpage
\section{Return forecasting}\label{a-alpha}

In this section we describe the return forecast
$\alpha_t$ used by the Markowitz portfolio, which is based on simple regression.

\subsection{Target and features}

\paragraph{Target and features.}
The forecast target is $r^\text{forw}_t \in \reals^3$, the $100$-trading day forward
return of the three assets, annualized by scaling by $252/100$.
It is computed from the same adjusted close prices used in the simulations
(\S\ref{ss-datasim}).
The feature vector $z_t \in \reals^p$ is described in
table~\ref{t-features}.
The asset-level features---trailing returns, volatilities, and trading
volumes---are computed from the ETF price and volume data; the rates and
market-state features are downloaded from FRED; and the factor features are
downloaded from the Kenneth French data library.
For forecasting on day $t$,
all features are computed using only data available on day $t$, on a rolling basis.
The total number of features is $p=42$.

\begin{table}
  \centering
  \small
  \begin{tabular}{lrp{0.65\linewidth}}
    \toprule
    Group & Number & Features \\
	    \midrule
    Asset trend & 12
      & Trailing returns of ETFs over $63$, $126$, and $252$ days,
      and the $252$-day trailing return divided by its trailing volatility.\\
    Asset volatility & 3
      & Log short ($21$ day) minus log long-run ($252$ day) volatility of ETFs.\\
    Asset liquidity & 9
      & Average log trading volume of ETFs over $21$, $63$, and $126$ days.\\
    Rates & 7
      & Treasury yields at $3$ months, $2$, $5$, $10$, and $30$ years,
      the $10$-year real yield, and the $10$-year breakeven inflation rate.\\
    Market state & 3
      & Smoothed ($63$ day average) log VIX, smoothed ($512$ day average)
broad dollar index, and the $2$-year minus $3$-month yield-curve slope.\\
    Fama--French & 8
      & Smoothed ($63$ day average) returns of market, size, value,
profitability, investment, momentum, short-term reversal, and
long-term reversal factors.\\
    \bottomrule
  \end{tabular}
  \caption{The feature vector $z_t$ used for the return forecast.}%
  \label{t-features}
\end{table}

\subsection{Forecasting method}
On each day $t$, we fit a multivariate ridge regression from the feature
vector $z_\tau$ to the forward-return vector $r^\text{forw}_\tau$.
The training set
contains at most the most recent $512$ complete observations. Its final target
observation is dated $t-100$, so that the corresponding forward return ends at
$t$ and no target uses prices after the forecast date, \ie, the method is causal.
We require at least $63$ complete observations; before then, the forecast is set to zero.

Within each training window, feature $j$ is divided by its sample (equal weight) standard
deviation $d_j$ (with a small numerical floor). Features are not centered, and
no intercept is included. Let $\tilde z_\tau \in \reals^p$ denote this scaled feature vector.
We choose the coefficient matrix $B_t \in \reals^{3\times p}$ by solving
the convex quadratic optimization problem
\begin{equation}\label{eq-ridge-alpha}
\begin{array}{ll} \mbox{minimize} &
\sum_{\tau} \rho_\tau
\left\|B \tilde z_\tau - r^\text{forw}_\tau\right\|_2^2
+ \lambda\|B\|_F^2,
\end{array}
\end{equation}
with variable $B \in \reals^{3\times p}$,
where the normalized observation weights $\rho_\tau$ decay exponentially with
a half-life of $252$ trading days.
We use regularization parameter value $\lambda=10$. The annualized return forecast is
then
\[
\alpha^{\rm ann}_t = B_t \tilde z_t \in \reals^3,
\]
where $B_t$ is the solution of \eqref{eq-ridge-alpha} and $\tilde z_t$ is the scaled feature vector on day $t$.
Only features observed throughout the training window and on day $t$ are used.
For the monthly portfolio decision in \eqref{eq-alphaopt}, we put this forecast
on the same $21$-trading-day scale as the cash return by using
$\alpha_t=(21/252)\alpha^{\rm ann}_t$.  (The simple Markowitz forecast is put on
the same scale by multiplying its exponentially weighted average daily return
by $21$.)
This simple forecast model is a ridge-regularized linear forecast,
re-fit each trading day using a rolling sample.

\subsection{Forecast performance}
\begin{figure}
    \centering
    \includegraphics[width=0.92\linewidth]{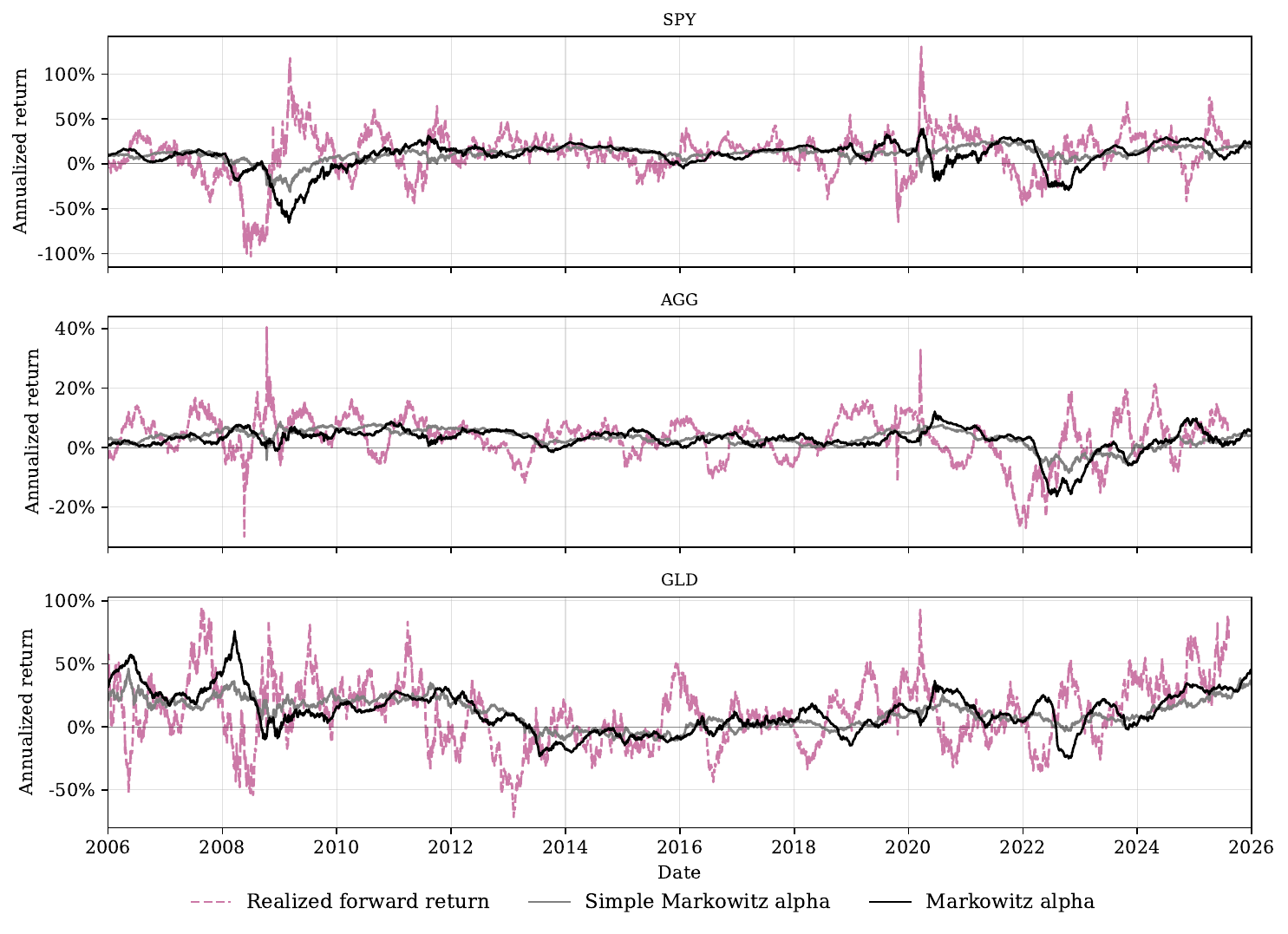}
    \caption{
        The two return forecasts and the realized $100$-day forward return
        $r^{\mathrm{forw}}_t$, plotted at the forecast date $t$, one panel per asset.
        The forecasts and realized returns are shown in annualized units,
        without normalization.
    }
    \label{f-alphas-sbg}
\end{figure}

The return forecasts used by the simple Markowitz and Markowitz
portfolios are shown in figure~\ref{f-alphas-sbg}, alongside the
realized $100$-day forward return (the target for the forecast method).
The trajectories are shown in their original
annualized units. The trailing average is smooth, while the
regression forecast responds to a broader set of market information.
The two forecasts are generally, but not always, in agreement.
It is easy to identify times when both forecasts are poor, \eg,
have the wrong sign compared to the realized 100-day forward returns.

\begin{figure}
    \centering
    \includegraphics[width=0.92\linewidth]{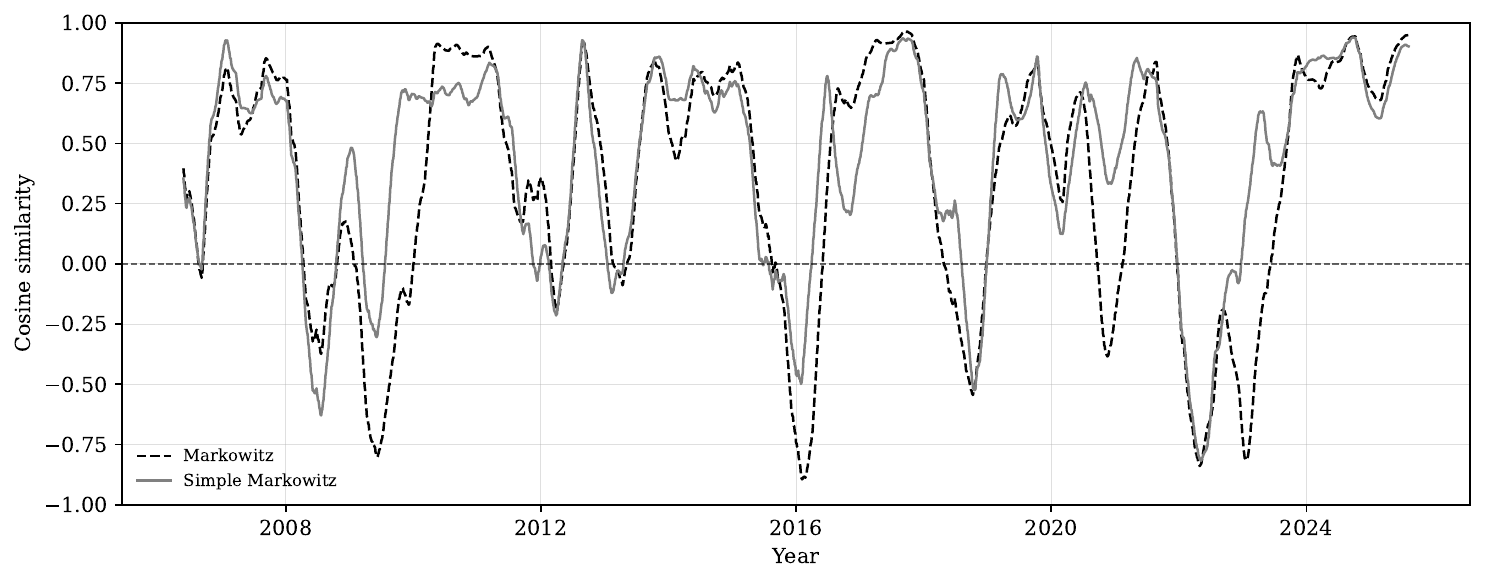}
    \caption{
        Rolling $100$-day mean of the cosine similarity between the return
        forecast $\alpha_t$ and the realized $100$-day forward return
        $r^{(h)}_t$.
    }
    \label{f-alpha-cosine-similarity}
\end{figure}

We evaluate the forecast performance by computing the cosine similarity
between the forecast at time $t$ and the realized $100$-day forward return at time $t$.
The results are shown in figure~\ref{f-alpha-cosine-similarity}.  The good news is that
these are generally positive.
From comparison with
figure~\ref{f-navs}, we see that the periods with large cosine
similarity correspond to periods with large cumulative returns.

\clearpage
\section{Statistical significance}\label{a-significance}

All confidence intervals reported here come from a stationary block bootstrap
\cite{Politis1994}, with a mean block length of
$21$ trading days and $10{,}000$ replications applied on the portfolio returns time series.
The stationary bootstrap preserves dependence within randomly selected runs of
consecutive observations. So if the time-series has a
short-memory autocorrelation, that structure is approximately retained in each replication.
Within each replication we sample the time series of returns for each portfolio (and the fed funds rate series) in the same manner
(dictated by the index series produced by the bootstrap),
so the contemporaneous correlation between the
portfolios is preserved, and the Sharpe ratio differences reported below are
paired.

Table~\ref{t-bootstrap} gives the point estimate and the $95\%$ interval for the
return, Sharpe ratio, and maximum drawdown of each of the six portfolios.

\begin{table}
  \centering
  \small
  \setlength{\tabcolsep}{3pt}
  \begin{tabular}{lrrrrrr}
    \toprule
    Portfolio & Return & 95\% interval & Sharpe & 95\% interval & Max DD & 95\% interval\\
    \midrule
    Markowitz        & 11.6\% & [7.5\%, 15.7\%]  & 1.08 & [0.62, 1.57] & 18.1\% & [9.9\%, 26.1\%] \\
    Simple Markowitz & 10.4\% & [6.4\%, 14.5\%]  & 0.91 & [0.48, 1.38] & 15.8\% & [10.5\%, 25.9\%] \\
    \midrule
    50/30/20 VC      &  7.8\% & [4.7\%, 10.8\%]  & 0.82 & [0.39, 1.26] & 15.9\% & [8.5\%, 22.4\%] \\
    60/40 VC         &  7.1\% & [4.0\%, 10.1\%]  & 0.71 & [0.29, 1.16] & 16.9\% & [9.1\%, 23.9\%] \\
    \midrule
    50/30/20         &  9.1\% & [4.7\%, 13.3\%]  & 0.70 & [0.27, 1.17] & 27.1\% & [12.7\%, 36.2\%] \\
    60/40            &  8.2\% & [3.4\%, 12.8\%]  & 0.56 & [0.13, 1.04] & 33.7\% & [14.9\%, 41.9\%] \\
    \bottomrule
  \end{tabular}
  \caption{Point estimates and bootstrap $95\%$ intervals for the return, Sharpe
ratio, and maximum drawdown of the six portfolios.}
  \label{t-bootstrap}
\end{table}

Table~\ref{t-sharpe-diff} gives the difference between the Sharpe ratio of the
Markowitz portfolio and that of each of the other five portfolios, with its
$95\%$ interval and the bootstrap probability that the difference is not
positive.

\begin{table}
  \centering
  \small
  \setlength{\tabcolsep}{3pt}
  \begin{tabular}{lrrr}
    \toprule
    Portfolio & Mean of $\Delta$ & 95\% interval & $P(\Delta \leq 0)$\\
    \midrule
    Simple Markowitz & $+0.17$ & $[-0.06, +0.40]$ & 0.068 \\
    \midrule
    50/30/20 VC      & $+0.26$ & $[-0.06, +0.61]$ & 0.057 \\
    60/40 VC         & $+0.37$ & $[-0.02, +0.78]$ & 0.033 \\
    \midrule
    50/30/20         & $+0.38$ & $[-0.01, +0.79]$ & 0.027 \\
    60/40            & $+0.52$ & $[+0.07, +0.99]$ & 0.011 \\
    \bottomrule
  \end{tabular}
  \caption{Difference $\Delta$ between the Sharpe ratio of the Markowitz
portfolio and that of each of the other portfolios, paired across bootstrap
replications, with its $95\%$ interval and the bootstrap probability that it is
not positive.}
  \label{t-sharpe-diff}
\end{table}

The intervals are wide, as is expected for Sharpe ratios estimated over twenty
years.
The advantage of the Markowitz portfolio over the 60/40 fixed-weight
benchmark---the principal benchmark this paper sets out to beat---is
significant at the $5\%$ level: its interval excludes zero.  The interval for
the 50/30/20 fixed-weight benchmark narrowly includes zero.

The advantages over the volatility-controlled benchmarks and over the simple
Markowitz portfolio are of comparable magnitude in point estimate---the
difference over the 60/40 VC benchmark, $+0.37$, is very nearly the difference
over the 50/30/20 benchmark, $+0.38$---but their intervals include zero, so they
are not significant at the $5\%$ level on this sample.
(The intervals are two-sided, so significance at this level corresponds to
$P(\Delta \leq 0) \leq 0.025$.)
Twenty years of data cannot resolve differences of this size.
These last comparisons are among portfolios we propose here: the
volatility-controlled benchmarks and the simple Markowitz portfolio are
constructions of this paper rather than established methods, so they bear on
which of our portfolios an investor should prefer, and not on whether we improve
on the status quo.
We therefore do not claim that the Markowitz portfolio is significantly better
than its volatility-controlled counterparts, only that it ranks above them
consistently.
We note that these intervals describe sampling variation only.
They do not account for the specification search discussed in
appendix~\ref{a-hyper-params}.

Table~\ref{t-subperiods} gives the Sharpe ratios of the six portfolios over the
four consecutive five-year subperiods, and over the full period.
The Markowitz portfolio has the highest Sharpe ratio in two of the four
subperiods, and over the full period.
In 2011--2016 it does not: its Sharpe is $0.86$, below the simple Markowitz
portfolio, at $0.94$, and both of the 60/40 portfolios, at $0.99$.
Its margin over the volatility-controlled benchmarks is widest in the first
subperiod, which contains the 2008 crisis, and is narrower afterwards.
Excluding the 2008 and 2009 calendar years entirely and splicing the remaining
daily returns, the Sharpe ratios are $1.11$ for the Markowitz portfolio, $0.95$
for the simple Markowitz portfolio, $0.93$ for 50/30/20 VC, $0.83$ for 60/40 VC,
$0.87$ for 50/30/20, and $0.74$ for 60/40, so the ordering is essentially
preserved---only 50/30/20 moves above 60/40 VC---but the margin narrows.

\subsection{End-to-end bootstrap}

As a more stringent check, we also apply the stationary bootstrap end to end,
using a mean block length of $300$ trading days and $100$ replications.  The
longer block preserves most of the dependence between the current features and
the $100$-day forward returns forecast by the regression.  Within
each replication, a common block index resamples the daily asset returns,
trading volumes, macro-financial series, Fama--French factors, and federal funds
rate.  We chain the sampled asset returns into continuous synthetic price
paths, reconstruct all features from these raw inputs, re-estimate the return
forecasts and covariance matrices, and rerun all six portfolio strategies.  The
common index preserves contemporaneous dependence across inputs and makes the
portfolio comparisons paired.

The results are given in table~\ref{t-bootstrap-end-to-end}. 
Every interval for the Markowitz
Sharpe advantage contains zero, and the estimated probability of a nonpositive
difference ranges from $0.28$ to $0.51$.  Even in this adverse evaluation setting, 
the Markowitz portfolio modestly outperforms the simple Markowitz, the $60/40$ and 
the $60/40$ volatility-controlled portfolios, and is essentially tied with the $50/30/20$ and
$50/30/20$ volatility-controlled portfolios.
With only $100$ replications, these
end-to-end results should be viewed as a conservative robustness check rather
than a precise tail-probability calculation.

\begin{table}
  \centering
  \small
  \setlength{\tabcolsep}{3pt}
  \begin{tabular}{lrrr}
    \toprule
    Portfolio & Mean of $\Delta$ & 95\% interval & $P(\Delta \leq 0)$\\
    \midrule
    Simple Markowitz & $+0.05$ & $[-0.16, +0.26]$ & 0.340 \\
    \midrule
    50/30/20 VC      & $-0.01$ & $[-0.29, +0.32]$ & 0.510 \\
    60/40 VC         & $+0.13$ & $[-0.22, +0.63]$ & 0.300 \\
    \midrule
    50/30/20         & $+0.02$ & $[-0.31, +0.34]$ & 0.480 \\
    60/40            & $+0.17$ & $[-0.23, +0.64]$ & 0.280 \\
    \bottomrule
  \end{tabular}
  \caption{Difference $\Delta$ between the Sharpe ratio of the Markowitz
  portfolio and that of each of the other portfolios in the end-to-end
  bootstrap with mean block length $300$ trading days, paired across $100$
  replications, with its $95\%$ interval and the bootstrap probability that it
  is not positive.}
  \label{t-bootstrap-end-to-end}
\end{table}

\clearpage
\section{Lagged information}\label{a-lag}

In this section we experiment with the effect on performance
of lagging the information provided
to the Markowitz portfolio by one day, one week, and one month.
In the first experiment we lag $\alpha_t$, but not $\Sigma_t$, with
the results shown in table~\ref{t-lagged-markowitz-alpha-only}.
We can see that lags in $\alpha_t$
barely affect the portfolio performance,
with only a modest degradation with full month lag.
In summary, our simple return forecast $\alpha_t$ is not sensitive to lag.

\begin{table}
  \centering
  \small
  \setlength{\tabcolsep}{3pt}
  \begin{tabular}{rrrrrrr}
    \toprule
    Lag & Return & Volatility & Sharpe & Max DD & Mean DD & Turnover \\
    \midrule
    0 days  & 11.6\% & 9.0\% & 1.08 & 18.1\% & 3.0\% & 284.4\% \\
    1 day   & 11.6\% & 9.0\% & 1.08 & 18.1\% & 2.9\% & 284.8\% \\
    5 days  & 11.5\% & 9.0\% & 1.07 & 17.6\% & 2.9\% & 285.3\% \\
    21 days & 10.9\% & 9.0\% & 1.00 & 15.1\% & 3.3\% & 284.2\% \\
    \bottomrule
  \end{tabular}
  \caption{Performance of the Markowitz portfolio when its return
  forecast is lagged.}
  \label{t-lagged-markowitz-alpha-only}
\end{table}

In our second experiment, we lag both $\alpha_t$ and $\Sigma_t$, with
the results shown in table~\ref{t-lagged-markowitz}.
Here too a one day lag has little effect, but longer lags do degrade performance
more than when $\alpha_t$ alone is lagged.
Our conclusion is that in our Markowitz portfolio, a reactive covariance forecast
appears to be substantially more important than a reactive return forecast.
This is consistent with the covariance estimate's short $11$-day window,
which is intended to respond quickly to changing market risk.

\begin{table}
  \centering
  \small
  \setlength{\tabcolsep}{3pt}
  \begin{tabular}{rrrrrrr}
    \toprule
    Lag & Return & Volatility & Sharpe & Max DD & Mean DD & Turnover \\
    \midrule
    0 days  & 11.6\% & 9.0\%  & 1.08 & 18.1\% & 3.0\% & 284.4\% \\
    1 day   & 10.9\% & 8.9\%  & 1.01 & 18.8\% & 2.8\% & 268.8\% \\
    5 days  & 10.1\% & 10.2\% & 0.81 & 30.6\% & 3.6\% & 274.5\% \\
    21 days &  9.4\% & 10.3\% & 0.73 & 23.0\% & 4.4\% & 285.4\% \\
    \bottomrule
  \end{tabular}
  \caption{Performance of the Markowitz portfolio when both its return
  forecast and covariance estimate are lagged.}
  \label{t-lagged-markowitz}
\end{table}

\clearpage
\section{Hyper-parameter sensitivity and choice}\label{a-hyper-params}

\paragraph{Sensitivity to backtest years.}
Our simulations and analyses are
for the specific span from 2006--2026, which includes several
market disruptions and multiple market conditions.  When other date ranges are
chosen, the performance numbers vary a bit, but the overall conclusions are almost always
the same: the volatility-controlled portfolios have better performance than the
fixed-weight benchmarks, and the Markowitz portfolios have even better performance. This is supported
by figure \ref{f-navs}.

\paragraph{Hyper-parameter sensitivity.}
We first address the question of sensitivity of our results for the Markowitz portfolio
to the hyper-parameters.
The Markowitz portfolio has very few hyper-parameters to choose. They are
target volatility (which we take as 7\%),
the allowed deviation of the relative weights from $(0.5,0.3,0.2)$
(which we take as one),
the trailing period for the covariance estimate (which we take as 11 days),
and the hyper-parameters in our return forecast.
Our choice of features, and their processing (listed in table~\ref{t-features})
seem to be unremarkable.
The hyper-parameters in the return forecast are the target horizon
(which we take as 100 days), the half-life used to fit the
regression matrix $B$ (which we take as 252 days), and the ridge regularization parameter
$\lambda$ (which we take to be $10$).

We analyze the performance for varying target volatilities in~\S\ref{ss-risktargets};
the results are what we would expect, and the high level conclusions remain.
The limit on relative weight deviation has a small effect on performance,
and could be removed without much change. The covariance estimate window does affect
the performance, though smoothly.
The three hyper-parameters in the return forecast have a smooth effect on performance.
As for choice of features, we can drop several without much decrease in performance.
(And we have no doubt that adding others could improve performance.)
In summary, the performance of the Markowitz portfolio is not particularly sensitive
to the choice of the (few) hyper-parameters.  Large changes do, however,
affect the performance.

\paragraph{Choice of hyper-parameters.}
Another question we address is how the Markowitz portfolio performance is affected depending on \emph{when} we choose the hyper-parameters,
by doing simulations on past data.
(As usual, hyper-parameters are crudely gridded.)
For a fixed 2006 start, the specification described in the main body (called current here) remains close to the best
tested one-at-a-time neighboring specification at every annual endpoint from
2011 through 2025 with respect to Sharpe ratio (calculated within each period).
At each endpoint, the current specification is well above the $60/40$ benchmark in terms of Sharpe ratio.
A longer alpha-fitting half-life of $315$ days performs
slightly better over these endpoints, but improves the full-period Sharpe ratio
by only $0.01$.

We also compare the fixed specification against annual walk-forward selection,
in which, at each year end, the specification is chosen by Sharpe ratio over a
trailing window, and is then
applied for the following (deployment) year.
The evaluation begins at the first year at which a selection can be made,
so the two selection rules are evaluated over different windows.
Table~\ref{t-walkforward} gives the results.
The trailing five-year rule sometimes chose neighboring covariance-window or
decay values.
Thus, annual retuning did not improve out-of-sample performance,
and the results appear locally robust, although shorter trailing windows do not
always select exactly the same parameters.

\begin{table}
  \centering
  \small
  \setlength{\tabcolsep}{3pt}
  \begin{tabular}{llrr}
    \toprule
    Selection rule & Evaluation window & Fixed specification & Walk-forward selection\\
    \midrule
    Trailing five years & 2011--2026 & 1.00 & 0.93 \\
    Trailing ten years  & 2016--2026 & 1.08 & 1.00 \\
    \bottomrule
  \end{tabular}
  \caption{Sharpe ratio of the fixed specification of the Markowitz portfolio (described in the main body) and of annual walk-forward
selection, for two selection windows.  The two rules are evaluated over
different windows, since the evaluation window begins at the first year at which
a selection can be made.}
  \label{t-walkforward}
\end{table}

\paragraph{Deflated Sharpe ratio.}
We account for the specification search using the deflated Sharpe ratio of
Bailey and L\'opez de Prado \cite{Bailey2014DeflatedSharpe, Bailey2017}.
The documented sweep contains $17$ specifications, whose annualized Sharpe
ratios have a standard deviation of $0.07$, which gives an expected maximum
Sharpe ratio, under the null hypothesis of no skill, of $0.12$.
The resulting deflated Sharpe ratio for the Markowitz portfolio exceeds $0.99$,
and remains above $0.99$ when the assumed number of trials is raised to $1000$,
which lifts the expected maximum under the null only to $0.22$.
This calculation assumes that the trials are the documented one-at-a-time
neighbors of the chosen specification.
These are highly correlated with one another, so the dispersion of the trial
Sharpe ratios---and hence the deflation---is likely understated.

\clearpage
\section{Comparison with risk-based portfolios}\label{a-risk-based}

Both our volatility-controlled and our optimization-based portfolios size their
positions using an estimate of risk, so it is natural to ask whether their
performance simply reflects a generic benefit of risk control, which is shared by
a large family of well known risk-based allocation rules.
In this appendix we address this question by comparing the Markowitz portfolio
against seven standard alternatives, implemented in our setting.

Each benchmark holds the same three assets (SPY, AGG, and GLD) plus cash, is
long only and unlevered, rebalances monthly, uses the same covariance estimate
$\Sigma_t$ described in \S\ref{ss-alphaport}, and pays the same
trading cost, over the same 2006--2026 period.
Each is evaluated using the metrics of \S\ref{ss-metrics}.
The comparison is therefore an apples-to-apples one:
the benchmarks differ from our portfolios only in how they turn the same
data into weights.

\subsection{Methods}\label{ss-risk-based-methods}
On each rebalance day, each of the seven methods produces a fully invested
vector of relative weights $\eta \in \reals^3$, satisfying $\eta \geq 0$ and
$\ones^T \eta = 1$.
Let $\nu_t \in \reals^3$ be the vector of estimated asset volatilities,
$(\nu_t)_i = (\Sigma_t)_{ii}^{1/2}$, and let
\[
\mathrm{RC}_i(\eta) = \frac{\eta_i (\Sigma_t \eta)_i}{\eta^T \Sigma_t \eta},
\quad i=1,2,3,
\]
be the fraction of portfolio variance contributed by asset $i$; these
fractions sum to one, though an individual fraction can be negative when the
covariance estimate has negative off-diagonal entries.
The seven methods are the following.
\begin{itemize}
\item \emph{Equal weight.} $\eta = (1/3)\ones$, the $1/N$ rule, which is a
surprisingly strong benchmark in practice \cite{DeMiguel2009}.
\item \emph{Inverse volatility.} $\eta_i \propto 1/(\nu_t)_i$, sometimes
called na\"{i}ve risk parity \cite{Qian2005, LeoteDeCarvalho2012}.
\item \emph{Equal risk contribution.} $\eta$ is chosen so that
$\mathrm{RC}_i(\eta) = 1/3$, \ie, each asset contributes the same share of
portfolio variance \cite{Maillard2010, Qian2005}.
\item \emph{50/30/20 risk budget.} $\eta$ is chosen so that
$\mathrm{RC}(\eta) = (0.5,0.3,0.2)$, \ie, \emph{risk}, rather than capital, is
split in the proportions of the 50/30/20 benchmark
\cite{Bruder2012, Roncalli2013}.
\item \emph{Global minimum variance.} $\eta$ minimizes $\eta^T \Sigma_t \eta$
\cite{Haugen1991, Clarke2006}.
\item \emph{Maximum diversification.} $\eta$ maximizes the diversification
ratio $(\nu_t^T \eta)/(\eta^T \Sigma_t \eta)^{1/2}$, \ie, the ratio of the
weighted average asset volatility to the portfolio volatility
\cite{Choueifaty2008}.
\item \emph{Black--Litterman.} The return forecast is shrunk toward the
returns implied by an equilibrium portfolio \cite{Black1992, He1999}.
We take the 50/30/20 benchmark $(w^{\text{tgt}} = (0.5, 0.3, 0.2))$ as the equilibrium portfolio, giving the implied
returns $\pi_t = \delta \Sigma_t w^\text{tgt}$ with risk aversion $\delta=2.5$,
and use the same return forecast $\alpha_t$ as our Markowitz portfolio
(appendix~\ref{a-alpha}) as absolute views on all three assets.
With prior covariance $\tau \Sigma_t$ and diagonal view covariance
$\Omega_t = \tau \diag ((\Sigma_t)_{11}, (\Sigma_t)_{22}, (\Sigma_t)_{33})$,
\ie, view uncertainty proportional to each asset's variance, and $\tau=0.05$,
the posterior mean return is
\[
\bar\mu_t = \left( (\tau \Sigma_t)^{-1} + \Omega_t^{-1} \right)^{-1}
\left( (\tau \Sigma_t)^{-1} \pi_t + \Omega_t^{-1} \alpha_t \right).
\]
The weights $\eta$ then maximize the mean-variance utility
$\bar\mu_t^T \eta - (\delta/2)\, \eta^T \Sigma_t \eta$.
(Here $\Sigma_t$ and $\alpha_t$ are expressed in annual units, since $\delta$
and $\tau$ are not scale free.)
\end{itemize}
The first six methods use only the covariance estimate $\Sigma_t$; only
Black--Litterman uses a forecast of returns, and it uses exactly the forecast
that our Markowitz portfolio uses.
The first two are given in closed form; the others are small smooth problems
over the simplex, which we solve numerically, warm started at the previous
month's solution.

None of these methods holds cash; each prescribes only the relative weights of
the three assets.  To compare them with our portfolios at a common level of
targeted risk, we also form a volatility-controlled version of each, exactly as
in \eqref{eq-vc}: on each rebalance day the weights are
\[
\hat w_t = \min \left\{ \frac{\sigma^\text{tgt}}
{(\eta^T \Sigma_t \eta)^{1/2}},\; 1 \right\} \eta ,
\]
with the same target $\sigma^\text{tgt}=7\%$ used throughout the paper, and the
remainder held in cash.  We report both the fully invested and the
volatility-controlled variants.

\subsection{Results}
Table~\ref{t-risk-based} gives the metrics for the seven benchmarks, in both
variants, with the Markowitz portfolio repeated from table~\ref{t-results} for
reference.  Table~\ref{t-risk-based-weights} gives their average relative
weights, along with the average cash weight of the volatility-controlled
variants.  (The relative weights do not depend on the variant, since scaling
toward the volatility target does not change them.)

\begin{table}
  \centering
  \small
  \setlength{\tabcolsep}{3pt}
  \begin{tabular}{lrrrrrr}
    \toprule
    Portfolio& Return & Volatility & Sharpe & Max DD & Mean DD &Turnover\\
    \midrule
    Markowitz                & 11.6\% &  9.0\% & 1.08 & 18.1\% & 3.0\% & 284.4\% \\
    \midrule
    \multicolumn{7}{l}{\emph{Volatility-controlled, 7\% target}}\\
    50/30/20 risk budget     &  6.6\% &  5.6\% & 0.85 & 14.7\% & 1.8\% & 178.9\% \\
    Equal risk contribution  &  6.2\% &  5.3\% & 0.83 & 14.7\% & 2.0\% & 158.5\% \\
    Equal weight             &  7.7\% &  7.2\% & 0.82 & 14.2\% & 2.6\% &  74.2\% \\
    Maximum diversification  &  6.6\% &  5.9\% & 0.82 & 15.8\% & 2.1\% & 276.3\% \\
    Inverse volatility       &  6.1\% &  5.4\% & 0.80 & 15.1\% & 2.0\% & 126.1\% \\
    Black--Litterman         &  8.5\% &  8.6\% & 0.77 & 16.2\% & 3.1\% & 270.4\% \\
    Global minimum variance  &  4.7\% &  4.9\% & 0.60 & 16.2\% & 1.8\% & 161.8\% \\
    \midrule
    \multicolumn{7}{l}{\emph{Fully invested}}\\
    50/30/20 risk budget     &  7.3\% &  7.1\% & 0.77 & 18.7\% & 1.9\% & 189.9\% \\
    Equal risk contribution  &  6.4\% &  6.9\% & 0.67 & 19.4\% & 2.0\% & 168.1\% \\
    Equal weight             &  8.7\% &  9.3\% & 0.73 & 23.8\% & 2.9\% &  13.5\% \\
    Maximum diversification  &  6.9\% &  6.7\% & 0.75 & 22.0\% & 2.3\% & 278.3\% \\
    Inverse volatility       &  6.6\% &  6.1\% & 0.80 & 17.0\% & 1.9\% & 120.7\% \\
    Black--Litterman         & 11.4\% & 15.2\% & 0.63 & 27.7\% & 6.1\% & 315.3\% \\
    Global minimum variance  &  5.1\% &  5.3\% & 0.63 & 17.1\% & 1.8\% & 157.8\% \\
    \bottomrule
  \end{tabular}
  \caption{Performance metrics for the seven risk-based benchmarks, in the
volatility-controlled (middle) and fully invested (bottom) variants, with the
Markowitz portfolio for reference (top).  Within each panel the methods are
ordered by the Sharpe ratio of the volatility-controlled variant.}
  \label{t-risk-based}
\end{table}

\begin{table}
  \centering
  \small
  \setlength{\tabcolsep}{3pt}
  \begin{tabular}{lrrrr}
    \toprule
    Portfolio & SPY & AGG & GLD & Cash\\
    \midrule
    Markowitz                & 48.3\% & 23.4\% & 28.3\% & 15.1\% \\
    \midrule
    50/30/20 risk budget     & 27.5\% & 58.7\% & 13.7\% &  5.4\% \\
    Equal risk contribution  & 20.5\% & 63.0\% & 16.5\% &  3.2\% \\
    Equal weight             & 33.4\% & 33.2\% & 33.4\% & 12.9\% \\
    Maximum diversification  & 23.0\% & 61.6\% & 15.4\% &  2.8\% \\
    Inverse volatility       & 18.7\% & 64.2\% & 17.1\% &  2.1\% \\
    Black--Litterman         & 44.9\% & 12.1\% & 43.0\% & 33.0\% \\
    Global minimum variance  & 10.8\% & 84.9\% &  4.4\% &  1.2\% \\
    \bottomrule
  \end{tabular}
  \caption{Average relative weights of the three assets, and average cash weight
of the volatility-controlled variant, for the seven risk-based benchmarks and
the Markowitz portfolio, averaged over all trading days in the evaluation
window.  The relative weights are the same for both variants.}
  \label{t-risk-based-weights}
\end{table}

\paragraph{Risk control alone does not explain the performance.}
At their standard specifications, every one of the seven methods lands well
below the Markowitz portfolio.
With volatility control, their Sharpe ratios range from $0.60$ to $0.85$,
against $1.08$ for the Markowitz portfolio.
The best of them, the 50/30/20 risk budget portfolio, at $0.85$, is only
slightly above our own 50/30/20 volatility-controlled benchmark, at $0.82$;
indeed the four best methods are, for practical purposes, tied with it.
The picture is much the same without volatility control, where the Sharpe ratios
range from $0.63$ to $0.80$.
It is worth stating this comparison the other way around.
In their conventional fully invested form, every one of the seven methods falls
below our 50/30/20 volatility-controlled benchmark; the highest of them,
inverse volatility, reaches $0.80$ against its $0.82$.
The four that do edge past it are those given our volatility-control overlay.
Sizing positions by risk, however it is done, gets an investor to roughly the
same place as the simple volatility-controlled benchmark of \S\ref{ss-vc},
and no further.

\paragraph{The comparison is not an artifact of the risk level.}
Because these methods cannot lever up, several of them realize volatilities well
below the $7\%$ target---as low as $4.9\%$ for the global minimum variance
portfolio---so we compare them on Sharpe ratio rather than return.
The Markowitz portfolio's Sharpe ratio is in any case not very sensitive
to the risk level (figure~\ref{f-pareto-sharpe}):
run at a $3\%$ target it realizes $4.8\%$ volatility with a Sharpe ratio
of $0.92$, and at a $5\%$ target it realizes $7.1\%$ volatility with a Sharpe
ratio of $1.07$.
So at each benchmark's own realized volatility, the Markowitz portfolio still
has a substantially higher Sharpe ratio.

\paragraph{Where the risk-based methods lose.}
Five of the six purely risk-based methods (all but equal weight) allocate using
the covariance estimate alone, and are therefore drawn toward the lowest
volatility asset.
As table~\ref{t-risk-based-weights} shows, they hold between $59\%$ and $85\%$
of their risky value in AGG, whose annualized return over the period is
$3.1\%$ (table~\ref{t-results}).
Their returns---$4.7\%$ to $6.6\%$ with volatility control, and $5.1\%$ to
$7.3\%$ fully invested---reflect this: they diversify risk well, but
they have no mechanism for tilting toward the assets that are expected to
perform well.
The Markowitz portfolio, by contrast, holds a far more balanced average mix,
$48.3/23.4/28.3$, and---as discussed in \S\ref{ss-weights-time}---derives
its performance from varying that mix over time rather than from its average.

\paragraph{The return forecast is necessary but not sufficient.}
The Black--Litterman portfolio uses the same return forecast $\alpha_t$ as our
Markowitz portfolio, and it does earn a higher return than the purely
risk-based methods, $8.5\%$ with volatility control and $11.4\%$ without.
But it earns those returns at a much higher risk: fully invested it realizes
$15.2\%$ volatility with a $27.7\%$ maximum drawdown, and even with volatility
control its Sharpe ratio is $0.77$.
Trading off risk against return in a utility function, with the forecast shrunk
toward an equilibrium prior, is simply less effective here than imposing a hard
constraint on the estimated volatility and an $\ell_1$ trust region around the
50/30/20 mix \eqref{e-ell_1-constr}, as in \eqref{eq-alphaopt}.
The lift we report therefore comes neither from the return forecast alone nor
from risk control alone, but from the combination.
Black--Litterman is the only benchmark considered here with free
hyper-parameters; we sweep them in \S\ref{ss-bl-sweep}, and its advantage over
the purely risk-based methods survives, but the ranking against the Markowitz
portfolio does not change.

\paragraph{Turnover.}
The Markowitz portfolio has the highest turnover of the portfolios in
table~\ref{t-risk-based}, at $284.4\%$.  But turnover does not explain the ranking: the maximum
diversification and Black--Litterman portfolios trade nearly as much,
at $276.3\%$ and $270.4\%$, with Sharpe ratios of $0.82$ and $0.77$; and all of
the figures reported here are net of the same trading costs.

\paragraph{The covariance estimate does not favor our method.}
All seven methods use the covariance estimate $\Sigma_t$ of
\S\ref{ss-alphaport}, whose $11$-day window we chose for our own portfolios, so
one might ask whether that choice handicaps them.
We therefore re-ran every method, in both variants, over a grid of trailing
covariance windows of $11$, $21$, $63$, $126$, and $252$ trading days, \ie,
$70$ specifications in all.
Table~\ref{t-cov-grid} gives the Sharpe ratio of the volatility-controlled
variant.

\begin{table}
  \centering
  \small
  \setlength{\tabcolsep}{3pt}
  \begin{tabular}{lrrrrr}
    \toprule
    Method & 11 days & 21 days & 63 days & 126 days & 252 days\\
    \midrule
    50/30/20 risk budget     & 0.85 & 0.75 & 0.73 & 0.68 & 0.68 \\
    Equal risk contribution  & 0.83 & 0.77 & 0.74 & 0.69 & 0.70 \\
    Equal weight             & 0.82 & 0.78 & 0.72 & 0.69 & 0.67 \\
    Maximum diversification  & 0.82 & 0.78 & 0.75 & 0.69 & 0.71 \\
    Inverse volatility       & 0.80 & 0.74 & 0.73 & 0.70 & 0.69 \\
    Black--Litterman         & 0.77 & 0.70 & 0.66 & 0.66 & 0.70 \\
    Global minimum variance  & 0.60 & 0.45 & 0.42 & 0.37 & 0.37 \\
    \bottomrule
  \end{tabular}
  \caption{Sharpe ratio of the volatility-controlled variant of each of the
seven risk-based benchmarks, for trailing covariance windows of $11$ to $252$
trading days.  The methods are ordered as in table~\ref{t-risk-based}.}
  \label{t-cov-grid}
\end{table}

For every one of the seven methods the volatility-controlled variant attains its
highest Sharpe ratio at the $11$-day window---the window our own portfolios
use---and longer, more conventional windows make every method worse,
monotonically or nearly so; the shared covariance estimate helps the
comparators, it does not handicap them.
Taking the best cell of the whole grid, selected in sample, the highest Sharpe
ratio any of these methods reaches is $0.85$, against $1.08$ for the Markowitz
portfolio; in the fully invested variant the best cell anywhere in the grid is
$0.80$, for inverse volatility at the $11$-day window.

\subsection{Tuning Black--Litterman}\label{ss-bl-sweep}
The six purely risk-based methods have no free hyper-parameters beyond the
covariance estimate, which they share with our portfolios.  Black--Litterman
does, so we sweep them and report its best in-sample performance.

We first note that the scale $\tau$ has no effect at all.
With prior covariance $\tau \Sigma_t$ and view covariance
$\Omega_t = \tau \diag (\Sigma_t)$, as in \S\ref{ss-risk-based-methods},
the factor $\tau$ cancels from the posterior mean, leaving
\[
\bar\mu_t = \left( \Sigma_t^{-1} + \diag (\Sigma_t)^{-1} \right)^{-1}
\left( \Sigma_t^{-1} \pi_t + \diag (\Sigma_t)^{-1} \alpha_t \right),
\]
where $\diag(\Sigma_t)$ denotes the diagonal part of $\Sigma_t$.
(In simulation, $\tau = 0.005$, $0.05$, and $0.5$ give identical results to six
significant figures.)
To vary the confidence placed in the views we therefore introduce a multiplier
$\omega > 0$, taking $\Omega_t = \omega \tau \diag (\Sigma_t)$, which gives
\[
\bar\mu_t = \left( \Sigma_t^{-1} + \omega^{-1}\diag (\Sigma_t)^{-1} \right)^{-1}
\left( \Sigma_t^{-1} \pi_t + \omega^{-1}\diag (\Sigma_t)^{-1} \alpha_t \right).
\]
As $\omega \to 0$ the posterior mean approaches the return forecast $\alpha_t$,
and as $\omega \to \infty$ it approaches the equilibrium returns $\pi_t$;
the canonical specification is $\omega = 1$.

We sweep the risk aversion $\delta$ over ten values from $0.5$ to $100$ and the
view confidence $\omega$ over nine values from $10^{-4}$ to $100$, giving $90$
specifications, each run in both variants.
Table~\ref{t-bl-sweep} reports the best of them, alongside the canonical
specification and the Markowitz portfolio.

\begin{table}
  \centering
  \small
  \setlength{\tabcolsep}{3pt}
  \begin{tabular}{lrrrrrr}
    \toprule
    Specification & Return & Volatility & Sharpe & Max DD & Mean DD &Turnover\\
    \midrule
    Markowitz & 11.6\% &  9.0\% & 1.08 & 18.1\% & 3.0\% & 284.4\% \\
    \midrule
    \multicolumn{7}{l}{\emph{Volatility-controlled, 7\% target}}\\
    Canonical, $\delta=2.5$, $\omega=1$ &  8.5\% &  8.6\% & 0.77 & 16.2\% & 3.1\% & 270.4\% \\
    Best, $\delta=20$, $\omega=10^{-4}$ & 10.1\% &  8.7\% & 0.95 & 19.1\% & 3.4\% & 327.4\% \\
    \midrule
    \multicolumn{7}{l}{\emph{Fully invested}}\\
    Canonical, $\delta=2.5$, $\omega=1$ & 11.4\% & 15.2\% & 0.63 & 27.7\% & 6.1\% & 315.3\% \\
    Best, $\delta=15$, $\omega=10^{-4}$ & 12.3\% & 11.0\% & 0.94 & 24.3\% & 4.4\% & 328.1\% \\
    \bottomrule
  \end{tabular}
  \caption{Black--Litterman under its canonical specification and under the best
of the $90$ specifications swept, selected in-sample on the Sharpe ratio, with
the Markowitz portfolio for reference.}
  \label{t-bl-sweep}
\end{table}

Three things are worth noting.
First, even at its best specification---selected with full hindsight over the
entire simulation, on the metric being compared---Black--Litterman reaches a
Sharpe ratio of $0.95$ with volatility control and $0.94$ fully invested,
against $1.08$ for the Markowitz portfolio.
Over all $90$ specifications the volatility-controlled Sharpe ratio ranges from
$0.70$ to $0.95$, and the fully invested one from $0.57$ to $0.94$;
all of them remain below $1.08$.
The comparison is a conservative one, since the hyper-parameters of the
Markowitz portfolio are not selected this way
(appendix~\ref{a-hyper-params}).

Second, the best specifications are those with $\omega$ near zero, \ie, those in
which the equilibrium prior is switched off entirely and the posterior mean is
just our return forecast $\alpha_t$.
The shrinkage that Black--Litterman is designed to provide does not help here;
what the portfolio gains, it gains from the forecast it is given.

Third, what remains once the prior is switched off is a difference in
formulation.  With $\omega \to 0$ the benchmark chooses a fully invested
portfolio maximizing $\alpha_t^T \eta - (\delta/2) \eta^T \Sigma_t \eta$ and
then scales it toward the volatility target, whereas our Markowitz portfolio
chooses the cash weight and the asset mix jointly, subject to a hard constraint
on estimated volatility and an $L_1$ trust region around 50/30/20.
That difference is worth about $0.13$ of Sharpe ratio here.

\subsection{Caveats}
These are implementations of the seven methods within the restrictive setting of
this paper: three assets, long only, no leverage, and monthly rebalancing.
Risk parity in particular is normally applied to a broader universe and with
leverage, which is how it reaches equity-like returns; denied leverage, it is
confined to the low-volatility corner of the universe.
Our claim is therefore the narrow one: for an investor working under the
constraints we consider, and using the same data, these methods do not reach the
performance of the Markowitz portfolio.

\clearpage
\section{Simulation assumptions}\label{a-assumptions}

\paragraph{Trading cost.}
Our simulations charge a spread of $5$ basis points on each trade, which we
consider conservative for the three highly liquid ETFs we use.
To check that our results do not rest on this, we re-run the six portfolios at
$10$ and $20$ basis points; table~\ref{t-cost-sensitivity} gives the Sharpe
ratios. At each spread, the optimization-based portfolios use that same spread
in their objective and the simulation charges it to realized value.
At four times the assumed cost the Markowitz portfolio still returns $11.1\%$,
at a Sharpe ratio of $1.03$, and the ranking of the six portfolios is unchanged
at every cost level.
The annually rebalanced benchmarks are essentially unaffected, since they
barely trade.

\begin{table}[ht]
  \centering
  \small
  \setlength{\tabcolsep}{3pt}
  \begin{tabular}{lrrr}
    \toprule
    Portfolio & 5 bp & 10 bp & 20 bp\\
    \midrule
    Markowitz        & 1.08 & 1.07 & 1.03 \\
    Simple Markowitz & 0.91 & 0.92 & 0.89 \\
    \midrule
    50/30/20 VC      & 0.82 & 0.81 & 0.80 \\
    60/40 VC         & 0.71 & 0.71 & 0.69 \\
    \midrule
    50/30/20         & 0.70 & 0.70 & 0.70 \\
    60/40            & 0.56 & 0.56 & 0.56 \\
    \bottomrule
  \end{tabular}
  \caption{Sharpe ratios of the six portfolios at trading cost rates of $5$, $10$,
and $20$ basis points.  The paper uses $5$ basis points throughout.}
  \label{t-cost-sensitivity}
\end{table}

\paragraph{The cash rate.}
Our simulations accrue cash at the effective federal funds rate, which an
individual investor cannot obtain directly; the practical substitute is a
Treasury bill fund, which yields slightly less.
We therefore re-run the simulations with cash accruing at the federal funds rate
less $25$ basis points annualized, while still measuring the Sharpe ratio in
excess of the federal funds rate itself.  The optimization-based portfolios also
use this lower cash return in their objective and therefore re-optimize their
allocations.  The simple Markowitz Sharpe ratio falls by $0.002$, and the
volatility-controlled portfolios fall by less than $0.007$.  The Markowitz
portfolio instead increases its risky allocation and its realized Sharpe ratio
rises by $0.014$ in this historical simulation; this is a reallocation effect,
not a direct benefit from receiving less interest.
The two fixed-weight benchmarks are unaffected, since they hold no cash.
The direct effect is small because even the Markowitz portfolio holds only around
$15\%$ of its value in cash on average.

\end{document}